\documentclass{svproc}
\usepackage{url}
\usepackage{epsfig,graphicx,color,subfigure}
\usepackage{amssymb,amsmath,amsfonts,verbatim,xkeyval,bm}
\usepackage{tikz}

\def\ggot{{\mathfrak g}}
\def\igot{{\mathfrak {i}}}

\def\Dgot{{\mathfrak D}}

\def\Pgot{{\mathfrak P}}

\def\vet#1{{\bm #1}}
\def\build#1_#2^#3{\mathrel{
\mathop{\kern 0pt#1}\limits_{#2}^{#3}}}
\def\reali{\mathbb{R}}
\def\complessi{\mathbb{C}}
\def\naturali{\mathbb{N}}
\def\interi{\mathbb{Z}}
\def\toro{\mathbb{T}}

\def\diff{{\rm d}}
\def\Ascr{\mathcal{A}}

\def\Cscr{\mathcal{C}}
\def\Dscr{\mathcal{D}}
\def\Escr{\mathcal{E}}
\def\Fscr{\mathcal{F}}

\def\Hscr{\mathcal{H}}
\def\Iscr{\mathcal{I}}
\def\Jscr{\mathcal{J}}
\def\Kscr{\mathcal{K}}
\def\Lscr{\mathcal{L}}

\def\Oscr{\mathcal{O}}
\def\Pscr{\mathcal{P}}
\def\Rscr{\mathcal{R}}
\def\Sscr{\mathcal{S}}

\def\Uscr{\mathcal{U}}

\def\epsilon{\varepsilon}
\def\rho{\varrho}

\def\phi{\varphi}
\def\theta{\vartheta}
\def\csi{\xi}
\def\imunit{{\rm i}}
\def\poisson#1#2{\left\{ #1,\,#2 \right\}}

\def\lie#1{\Lscr_{#1}}
\def\Lie#1{\Lscr_{#1}}
\def\Pset{{\Pgot}}
\def\PPset{\widehat{\Pgot}}
\def\fastpoisson#1#2{\left\{ #1,#2 \right\}_{\vet{L},\vet{\lambda}}}
\def\secpoisson#1#2{\left\{ #1,#2 \right\}_{\vet{\csi},\vet{\eta}}}
\def\wideitem#1{\par\hangindent\itemindent
   \noindent\hbox to\parindent{\hfil{#1}\enspace}\ignorespaces}
\def\biwideitem#1{\par\hangindent 3.\itemindent
   \noindent\hbox to 3.\parindent{\hfil{#1}\enspace}\ignorespaces}
\def\scalprod#1#2{#1\cdot#2}

\begin{document}
\mainmatter              
\title{Invariant KAM tori: from theory to applications to exoplanetary systems}
\titlerunning{Invariant KAM tori $\ldots$ applications to exoplanetary systems}  
%
\author{Ugo Locatelli\inst{1} \and Chiara Caracciolo\inst{2}
\and Marco Sansottera\inst{2} \and Mara Volpi\inst{1}}
\authorrunning{Ugo Locatelli et al.} 
%
\tocauthor{Ugo Locatelli, Chiara Caracciolo,
Marco Sansottera, Mara Volpi}
\institute{Dipartimento di Matematica dell'Universit\`a degli Studi di
  Roma ``Tor Vergata'', via della ricerca scientifica 1, 00133 Roma,
  Italy \and Dipartimento di Matematica dell'Universit\`a degli Studi
  di Milano, via Saldini 50, 20133 Milano, Italy}

\maketitle              

\begin{abstract}
  We consider the classical problem of the construction of invariant
  tori exploiting suitable Hamiltonian normal forms.  This kind of
  approach can be translated by means of the Lie series method into
  explicit computational algorithms, which are particularly suitable
  for applications in the field of Celestial Mechanics.  First, the
  algorithm constructing the Kolmogorov normal form is described in
  detail. Then, the extension to lower-dimensional elliptic tori is
  provided.  We adopt the same formalism and notations in both cases,
  with the aim of making the latter easier to understand. Finally,
  they are both used in a combined way in order to approximate
  carefully the secular dynamics of the extrasolar system hosting two
  planets orbiting around the HD~$4732$ star.
\end{abstract}

\section{Introduction}\label{sec:intro}
The birth of KAM\footnote{It is worth to repeat, here and once again,
  the story explaining the choice of the acronym KAM. In 1954, during
  the International Congress of Mathematicians in Amsterdam,
  Kolmogorov presented his version of the (KAM) theorem. In the same
  year, he also wrote the very short article~\cite{Kolmogorov-1954},
  where he provided just a scheme of the proof. According to a few
  direct witnesses, a few years later Kolmogorov explained all the
  details of his proof in a cycle of lectures delivered at the Moscow
  University. This was based on a sequence of canonical
  transformations coherently defined on a so called scale of Banach
  spaces; a modern reformulation of the proof that should be very
  similar to the original one is included in~\cite{Chierchia-2008}. In
  1963, V.I.~Arnold (who had been a student of Kolmogorov) published a
  complete proof of the theorem, based on a different approach able to
  ensure the existence of a Cantor set including many invariant tori
  and having positive Lebesgue measure (see the statement of
  Corollary~\ref{thm:KAM-Arnold} and~\cite{Arnold-1963}). In the
  meantime, the german mathematician J.~Moser developed a completely
  independent version of the proof in the case of symplectic mappings
  (see~\cite{Moser-1962}). Let us also recall that at the beginning
  the correctness of the Kolmogorov's approach was doubtful for
  Moser. Indeed, also because of a famous sentence included in the
  report he wrote for {\it Mathematical Reviews} on the Kolmogorov's
  article (see~MR0097508, 20 n. 4066), for many years Arnold's
  approach was thought to be the only viable one, in order to prove
  KAM theorem for quasi-integrable Hamiltonian systems.} theory was
marked by a famous article written in 1954 by A.N.~Kolmogorov,
i.e.,~\cite{Kolmogorov-1954}. At that epoch the great potential of KAM
theorem in order to solve interesting problems in the field of
Celestial Mechanics was immediately understood. In fact, it was
applied just a few years later in order to prove the stability of the
triangular Lagrangian points in the planar restricted problem of three
bodies for almost all admissible mass ratios
(see~\cite{Leontovich-62}).  Since then, several applications have
ensured the existence of invariant tori in the context of other
Hamiltonian models that are of interest in Celestial Mechanics (see,
e.g.,~\cite{Cel-Chi-2007}). Nevertheless, the applications of KAM
theory to physically realistic models have never been
straightforward. This is mainly due to a few severe constraints that
appear in the hypotheses of KAM theorem (e.g., concerning the
smallness on the parameter ruling the size of the perturbation).

In the last few decades, the successful applications of KAM theory to
Celestial Mechanics introduced more and more refinements in the
preliminary work to adapt the Hamiltonian model in such a way to
bypass the aforementioned difficulties (see, e.g.,~\cite{Loc-Gio-2000}
and~\cite{Loc-Gio-2007}). In some other works, the novelty concerns
the design of a new approach strategy. In particular, this has been
made by combining the results provided by two different theorems; for
instance, in~\cite{Gio-Loc-San-2009} and~\cite{Gio-Loc-San-2017} the
estimates {\it \`a la} Nekhoroshev have been applied in the
neighborhood of an invariant KAM torus, by following the proof scheme
described in~\cite{Mor-Gio-1995}. This kind of strategy can be
implemented in a natural way by adopting an approach based on suitable
\emph{normal forms}. Indeed, different normal form algorithms can be
applied one after the other. This work has the ambitious goal of fully
explaining a very recent type of applications in the field of
Celestial Mechanics, where the computational procedure leading to the
Kolmogorov normal form is performed in the neighborhood of a periodic
orbit. In turn, such an invariant manifold is preliminarly located by
a corresponding normal form for an elliptic torus. The addition of
this intermediate step is crucial in order to successfully apply our
computational algorithm in its entirety to extrasolar planetary
systems with rather eccentric orbits (i.e., whose eccentricity values
are significantly larger than those observed for the gaseous planets
of our Solar System).

The first theoretical results about the existence of elliptic tori go
back to~\cite{Melnikov-1965}, \cite{Eliasson-1988}
and~\cite{Poschel-1989}.  In the last two decades, similar statements
have been proved also in the context of Hamiltonian planetary systems
(see~\cite{Bia-Chi-Val-2003}, \cite{Bia-Chi-Val-2006}
and~\cite{Gio-Loc-San-2014}). In the present notes, we aim to develop
an approach that is far from being purely theoretical. Indeed, we will
explain how to extract from the proof schemes the information that is
fundamental in order to properly design a computational procedure,
which allows to determine invariant manifolds that are in good
agreement with the orbital motions of extrasolar planets.

In the following, Section~\ref{sec:KAMbasics} contains a quick
  introduction of a few elementary notions concerning the Hamiltonian
  perturbation theory and a careful description of the normal form
  method constructing KAM tori. In Section~\ref{sec:kolmog}, we show
  how that approach can be adapted for the construction of
  lower-dimensional invariant manifolds of elliptic type. In the
final Section~\ref{sec:exotori} our new application to an exoplanetary
system is explained in detail; this is designed by combining the two
kind of normal forms previously discussed, whose constructions are
performed one after each other.


\section{Basics of KAM theory}\label{sec:KAMbasics}

\subsection{Near to the identity canonical transformations by Lie series}\label{sbs:def-Lie-series}
Let us consider two generic dynamical functions $f=f(\vet{p},\vet{q})$
and $\chi=\chi(\vet{p},\vet{q})$, that are defined on all the phase
space endowed by $n$ pairs of conjugate canonical variables
$(\vet{p},\vet{q})=(p_1\,,\,\ldots\,,\,p_n\,,\,q_1\,,\,\ldots\,,\,q_n)$.
It is well known that the time evolution of $f$
under the flow induced by $\chi$ is ruled by the Poisson bracket between
these two functions, i.e.,
$\dot f=\frac{\diff}{\diff t}f(\vet{p}(t),\vet{q}(t))=
\{f,\chi\}$, where
\begin{equation}
  \label{def:Poisson-bracket}
  \{f\,,\,\chi\}=\sum_{j=1}^{n}
  \frac{\partial f}{\partial q_j}\frac{\partial \chi}{\partial p_j}-
  \frac{\partial f}{\partial p_j}\frac{\partial \chi}{\partial q_j}
\end{equation}
and the flow
$(\vet{p}(t),\vet{q}(t))=\Phi_\chi^t\big(\vet{p}(0),\vet{q}(0)\big)$
is defined by the solution of the corresponding Hamilton equations
\begin{equation}
  \label{def:Ham-eq-4-chi}
  \dot p_j = - \frac{\partial \chi}{\partial q_j}\ ,
  \qquad
  \dot q_j = \frac{\partial \chi}{\partial p_j}\ ,
  \qquad
  \forall\ j=1,\ldots,n
\end{equation}
(being $(\vet{p}(0),\vet{q}(0))$ regarded as initial conditions).

Let us now focus on the Taylor expansion with respect to time of the
generic dynamical function $f$, i.e., $f +t\dot
f+\frac{t^2}{2}\frac{\diff}{\diff t}{\dot f}+\ldots=
f+t\{f,\chi\}+\frac{t^2}{2}\{\{f,\chi\},\chi\}+\ldots$, that can be
reformulated in terms of Lie series. First, let us introduce the so
called Lie derivative operator: $\Lie{\chi}\cdot=\{\cdot,\chi\}$; {
  in the present context, it is usual to refer to $\chi$ as the
  generating function of the corresponding Lie derivative. Thus,} the
previous Taylor expansion in time can be expressed as
$\exp\big(t\Lie{\chi}\big)
f=\sum_{j=0}^\infty\frac{t^j}{j!}\Lie{\chi}^jf\,$.  It is common to
define the Lie series operator just in the case with $t=1$, i.e., {
  it acts on the generic dynamical function $f$ in such a way that}
$$
\exp\big(\Lie{\chi}\big)f=\sum_{j=0}^\infty\frac{1}{j!}\Lie{\chi}^j f\ ;
$$
let us emphasize that this formula must be interpreted at a purely
formal level, in the sense that we do not wonder about the convergence
of the series. However, it can be ensured if the sup-norm of the
generating function $\chi$ is small enough\footnote{The convergence of
  the Lie series is carefully discussed { in~\cite{Grobner-60}
    and~\cite{Giorgilli-2003}; in particular, the explanatory notes
    in~\cite{Giorgilli-2003}} contains also a rather self-consistent
  introduction to the Lie series formalism in the Hamiltonian
  framework.}, as it is natural to expect.

Since any single canonical coordinate can be seen as a particular
dynamical function, we can express the Hamiltonian flow at time~1 in
terms of Lie series in the following way:
\begin{equation}
  \label{def:Ham-flow-by-Lie_series}
  \Phi_\chi^1(\vet{p},\vet{q})=
  \exp\big(\Lie{\chi}\big)(\vet{p},\vet{q}) ,
\end{equation}
where, for every pair of canonical variables $(p_i\,,\,q_i)$ (being
$i=1,\ldots,n$), we put
$$
\Phi_\chi^1 p_{i}=\exp\big(\Lie{\chi}\big)p_{i}\ ,
\qquad
\Phi_\chi^1 q_{i}=\exp\big(\Lie{\chi}\big)q_{i}\ .
$$
It is well known that the Hamiltonian flow is canonical, then we
readily obtain that the map defined by the Lie series operator in the
right hand side of~\eqref{def:Ham-flow-by-Lie_series} is canonical as well.
Moreover, such a change of coordinates is obviously close to the
identity in the limit of the generating functions shrinking to zero.

The canonical formalism makes very convenient the writing of the
equations of motion in the new variables. Let us assume that the
evolution in the original set of coordinates $(\vet{p},\vet{q})$ is
ruled by a single function $H=H(\vet{p},\vet{q})$ entering the
Hamilton equations~\eqref{def:Ham-eq-4-chi} in place of $\chi$;
moreover, let $(\vet{p},\vet{q})=\Cscr(\vet{P},\vet{Q})$ be a
canonical transformation. Therefore, the new equations of motions can
be written as follows:
\begin{equation}
  \label{def:Ham-eq-4-Kscr}
  \dot P_j = - \frac{\partial \Kscr}{\partial Q_j}\ ,
  \qquad
  \dot Q_j = \frac{\partial \Kscr}{\partial P_j}\ ,
  \qquad
  \forall\ j=1,\ldots,n,
\end{equation}
being $\Kscr(\vet{P},\vet{Q})=H\big(\Cscr(\vet{P},\vet{Q})\big)$ the new
Hamiltonian function. In such a context, the Lie series formalism
makes automatic (and, then, somehow easier) the procedure of
substitution, because of the so called ``exchange theorem''
(see~\cite{Grobner-60}). In fact, if $\chi$ is a small enough
generating function, the new Hamiltonian can be expressed as
$$
\Kscr(\vet{P},\vet{Q}) =
\exp\big(\Lie{\chi}\big) H \Big|_{({\vet p},{\vet q}) = ({\vet P},{\vet Q})}\ ,
$$
this means that we can apply the Lie series to the old Hamiltonian
function so as to rename the variables, only at the end. For more
detailed explanations we defer to the whole Section~4.1
of~\cite{Giorgilli-2003}. Of course, the same computational procedure
holds also for the corresponding canonical transformation, that is given 
by
$$
(\vet{p},\vet{q}) = \Cscr(\vet{P},\vet{Q}) =
\exp\big(\Lie{\chi}\big) (\vet{p},\vet{q})
\Big|_{({\vet p},{\vet q}) = ({\vet P},{\vet Q})}\ .
$$


\subsection{Statement(s) of KAM theorem}\label{sbs:KAM-statements}
First, let us recall the statement of KAM theorem as in its very first
version introduced by Kolmogorov (see~\cite{Kolmogorov-1954}).
\begin{theorem} [KAM, according to the version due to Kolmogorov] \label{thm:KAM-Kolmogorov}
Consider a Hamiltonian function $H:\Ascr\times\toro^n\mapsto\reali$
(being $\Ascr\subseteq\reali^n$ an open set) of the form
$H(\vet{p},\vet{q})=\vet{\omega}\cdot\vet{p}+ h(\vet{p})+\epsilon
f(\vet{p},\vet{q})$ where $h$ is at least quadratic with respect to
the actions $\vet{p}$, i.e., $h(\vet{p})=\Oscr(\|\vet{p}\|^2)$ for
$\vet{p}\to\vet{0}$.  Moreover, let us assume the following
hypotheses:
\begin{itemize}
  \item[(a)] $\vet{\omega}$ is Diophantine; this means that there are
    two positive constants\footnote{Indeed, in order to satisfy the
    Diophantine inequality, it is essential that $\tau\ge n-1$.}
    $\gamma$ and $\tau$ such that $|\vet{k}\cdot\vet{\omega}|\ge
    \frac{\gamma}{|\vet{k}|^{\tau}}$
    $\forall\ \vet{k}\in\interi^n\setminus\{\vet{0}\}$;
  \item[(b)] $H$ is analytic on its action--angle\footnote{{
      Although there exist formulations of the KAM theorem that are
      not dealing with action--angle canonical coordinates (see,
      e.g.,~\cite{delaLlave-et-al-2005}), we stress that this is
      a rather natural framework to assume. In fact, by definition a
      $n$--dimensional torus $\toro^n$ is in a bijective
      correspondence with $n$ angles, denoted as
      $(q_1\,,\,\ldots\,,\,q_n)$ in agreement with the text.  Thus,
      they can be adopted as coordinates. Let us recall that in
      Hamiltonian mechanics the product between each conjugate pair of
      canonical variables has the physical dimension of an action,
      that is the same as an angular momentum. Therefore,
      $\forall\ j=1,\,\ldots\,n$, the conjugate momentum $p_j$ is an
      action, because $(\vet{p},\vet{q})$ are assumed to be canonical
      coordinates.}} domain of definition $\Ascr\times\toro^n$;
  \item[(c)] $h(\vet{p})$ is non-degenerate, i.e.,
  $\det\big(\frac{\partial^2\,h} {\partial p_i\partial p_j}
  (\vet{p})\big)_{i,j}\neq 0$ $\forall\ \vet{p}\in\Ascr$;
  \item[(d)] $\epsilon$ is a small enough parameter.
\end{itemize}
Therefore, there is a canonical transformation
$(\vet{p},\vet{q})=\psi_\epsilon(\vet{P},\vet{Q})$, leading $H$ in the
so called Kolmogorov normal form
$\Kscr(\vet{P},\vet{Q})=\omega\cdot\vet{P}+\Oscr(\|\vet{P}\|^2)$,
being $\Kscr=H\circ\psi_\epsilon\,$.
\end{theorem}

In our exposition of these topics, we do not consider all the very
interesting mathematical work that has been done in the last fifty
years in order to weaken the assumptions on the KAM theorem. We prefer
to focus on what makes the theorem suitable to apply to interesting
physical problems. This is somehow hidden in the thesis of the statement
and, mainly, in the proof scheme. Let us highlight such a content.

One can easily verify that, if the Hamiltonian is in the Kolmogorov
normal form
$\Kscr(\vet{P},\vet{Q})=\omega\cdot\vet{P}+\Oscr(\|\vet{P}\|^2)$, then
$t\mapsto(\vet{P}(t)=\vet{0}\,,\,\vet{Q}(t)=\vet{Q}_0+\vet{\omega}t)$
is the solution for the equations of motion~\eqref{def:Ham-eq-4-Kscr}
starting from the generic initial conditions
$(\vet{P}(0)\,,\,\vet{Q}(0))=(\vet{0}\,,\,\vet{Q}_0)$. Since the
canonical transformations enjoy the property of preserving solutions,
this allows us to design the following integration scheme for the
equations of motion~\eqref{def:Ham-flow-by-Lie_series}, when the generic
Hamiltonian $\chi$ is replaced by $H$, that describes the problem we
are considering:
\begin{equation}
  \label{frm:semi-anal-scheme}
  \vcenter{\openup1\jot\halign{
      \hbox to 20 ex{\hfil $\displaystyle {#}$\hfil}
      &\hbox to 14 ex{\hfil $\displaystyle {#}$\hfil}
      &\hbox to 20 ex{\hfil $\displaystyle {#}$\hfil}\cr
      \big(\boldsymbol{p}(0),\boldsymbol{q}(0)\big)
      &\build{\longrightarrow}_{}^{{{\displaystyle\psi_\epsilon^{-1}}
          \atop \phantom{0}}}
      &\left({{\displaystyle \boldsymbol{P}(0)}
        \,,\, {\displaystyle \boldsymbol{Q}(0)}}\right)
      \cr
      & &\Big\downarrow \build{\Phi_{\Kscr}^{t}}_{}^{}
      \cr
      \big(\boldsymbol{p}(t),\boldsymbol{q}(t)\big)
      &\build{\longleftarrow}_{}^{{{\displaystyle \psi_\epsilon} \atop \phantom{0}}}
      &\left({{\displaystyle  \boldsymbol{P}(t)}\,,\,
        {\displaystyle \boldsymbol{Q}(t)}}
      \right)
      \cr
  }}
  \ \ \ .
\end{equation}
In the scientific literature, this way to compute the motion law
$t\mapsto(\vet{p}(t)\,,\,\vet{q}(t))=
\Phi_{H}^{t}(\vet{p}(0)\,,\,\vet{q}(0))$ is often said to be
semi-analytic. Such a name is due to the fact that the schematic
procedure above is usually performed after having determined the
Fourier expansions of the canonical transformation $\psi_\epsilon\,$,
by using a software package designed for doing computer algebra
manipulations.

In spite of the fact that the very first version of the KAM theorem
ensures the existence of a single invariant torus, the statement can
be extended so as to cover a very generic situation. Indeed, in his
very short but incredibly seminal article~\cite{Kolmogorov-1954},
Kolmogorov recalled a well known result of number theory: almost all
$n$--dimensional vectors are Diophantine. This remark jointly with the
uniform non-degeneracy of the so called action-frequency map in the
integrable approximation, i.e., $\vet{p}\mapsto
\vet{\omega}(\vet{p})=\big(\frac{\partial\,h} {\partial
  p_i}(\vet{p})\big)_{i=1,\ldots,n}\,$, allowed him to state the
following result in~\cite{Kolmogorov-1954}.

\begin{corollary} [KAM, according to the version proved by Arnold] \label{thm:KAM-Arnold}
Consider a quasi-integrable Hamiltonian depending on action--angle
variables, i.e., $H:\Ascr\times\toro^n\mapsto\reali$ (being
$\Ascr\subseteq\reali^n$ an open set) of the form
$H(\vet{p},\vet{q})=h(\vet{p})+\epsilon f(\vet{p},\vet{q})$. If we
assume the same hypotheses~(b)--(d) of
Theorem~\ref{thm:KAM-Kolmogorov}, then there is a set $\Sscr_\epsilon$
that is made by invariant tori and is such that its Lebesgue measure
$\mu\big(\Sscr_\epsilon\big)$ is positive. Moreover,
$$
\lim_{\epsilon\to 0}
\mu\Big(\big(\Ascr\times\toro^n\big)\setminus\Sscr_\epsilon\Big)=0\ .
$$
\end{corollary}
Let us emphasize that this statement highlights one of the main merits
of the KAM theorem: it shows that there is a sort of continuity (in
terms of the Lebesgue measure) between integrable systems and
quasi-integrable ones. From one hand, this sort of intuitive
concept was (and still is) considered to be extremely natural; on
the other hand, at that epoch such an expectation was in contrast with
the famous theorem by Poincar\'e (that can be felt as somehow
paradoxical, see~\cite{Poincare-1892}) on the non-existence of
integrals of motion apart from the energy for a generic quasi-integrable
Hamiltonian system.

Although the statement of the Corollary above can be easily deduced
from the original version of the KAM theorem that is due to
Kolmogorov, the proof scheme introduced by Arnold
in~\cite{Arnold-1963} is extremely deep, because it provides a more
global picture of the dynamics. This approach has been further
extended, for instance, in~\cite{Poschel-82}, where it is proved that
quasi-integrable Hamiltonian satisfying the usual hypotheses~(b)--(d)
of Theorem~\ref{thm:KAM-Kolmogorov} can be conjugated to integrable
ones via a canonical transformation that is not analytic, but it is
$\Cscr^{(\infty)}$.

\subsection{Algorithmic construction of the Kolmogorov normal form}\label{sbs:Kolmog-original}

These notes are focusing more on the applications based on the KAM
theory rather than on the theory itself. Therefore, it is important to
describe carefully the so called formal algorithm constructing the
Kolmogorov normal form. The results about the convergence of such a
computational procedure are very well established { (see,
  e.g.,~\cite{Gio-Loc-1997})} and in the following we will just
briefly recall them.

For the sake of definiteness, we need to introduce some notations.
For a fixed positive integer $K$ we introduce the distinct
classes of functions $\Pscr_{\ell,sK}\,$, for all non-negative
indexes $\ell,\,s\ge 0\,$. Any generic function
$g\in\Pscr_{\ell,sK}$ can be written as
\begin{equation}
g( \vet p, \vet q) =
\sum_{{\scriptstyle{\vet j\in\naturali^{n}}}\atop{\scriptstyle{|\vet j|=\ell}}}
\,\sum_{{\scriptstyle{{ \vet k\in\interi^{n}}}\atop{\scriptstyle{| \vet k|\le sK}}}} 
\,c_{\vet j,\vet k} \vet p^{\vet j}\exp\big(\imunit\scalprod{\vet k}{ \vet q}\big)
\ ,
\label{frm:esempio-g-in-Pscr}
\end{equation}
where $( \vet p, \vet q)$ are action--angle canonical variables and
the coefficients $c_{\vet j,\vet k}\in\complessi$ satisfy the
following relation: $c_{\vet j,-\vet k}={\bar c}_{\vet j,\vet k}$ so
that $g:\,\reali^n\times\toro^n\mapsto\reali$. Moreover, in the
previous formula, we have introduced the symbol $|\cdot|$ to denote
the $\ell_1$-norm (i.e., $|\vet{k}|=|k_1|+\ldots+|k_n|$) and we have
adopted the multi-index notation, i.e.,
$\vet{p}^{\vet{j}}=p_1^{j_1}\cdot\ldots\cdot p_n^{j_n}$. In the
following, we will adopt the usual notation for the average of a
function $g$ with respect to the generic angles $\vet
\vartheta\in\toro^n$, i.e., $\langle g\rangle_{\vet \vartheta} =
\int_{\toro^n}\diff\vartheta_1\ldots\diff\vartheta_n \,g/(2\pi)^n$.

We will start the formal algorithm from a Hamiltonian
of the following type:
\begin{equation}
\label{frm:espansione-HKAM^(0)}
\vcenter{\openup1\jot 
\halign{
$\displaystyle\hfil#$&$\displaystyle{}#\hfil$&$\displaystyle#\hfil$\cr
  H^{(0)}(\vet p, \vet q;\vet \omega^{(0)})
  & \,=\, E^{(0)}+\scalprod {\vet \omega^{(0)}}{\vet p }+ 
  \sum_{s\ge 0}\sum_{\ell\ge 2} f_{\ell}^{(0,\,s)}(\vet p, \vet q;\vet \omega^{(0)})
  \cr
  &\,+\,
  \sum_{s\ge 1}\sum_{\ell =0}^1 f_{\ell}^{(0,\,s)}(\vet p, \vet q;\vet \omega^{(0)})
  \,,
  \cr
}}
\end{equation}
where $f_{\ell}^{(0,\,s)}\in \Pscr_{\ell,sK}$, being the first upper
index related to the normalization step, and $E^{(0)}\in \Pscr_{0,0}$
is a constant meaning the energy level of the torus $\big\{( \vet p,
\vet q)\,:\ \vet{p}=\vet{0},\ \vet{q}\in\toro^{n}\big\}\,$ that is
invariant in the integrable approximation. The occurrence of $\vet
\omega^{(0)}$ at the end of the list of the arguments emphasizes
that those functions depend also on that angular velocity
vector in a parametric way. We also stress that the terms appearing in
the second row of formula~\eqref{frm:espansione-HKAM^(0)} have to be
considered as the small perturbation we aim to remove in order to
bring the Hamiltonian in Kolmogorov normal form. According to the
definition given by Poincar\'e (see~\cite{Poincare-1892}), the
\emph{general problem of the dynamics} is described by a real analytic
Hamiltonian of type
$H(\vet{I},\vet{\phi};\epsilon)=h(\vet{I})+\epsilon
f(\vet{I},\vet{\phi})$, being $(\vet{I},\vet{\phi})$ action--angle
coordinates and $\epsilon$ a small parameter. It is well known that
such an Hamiltonian can be put in the
form~\eqref{frm:espansione-HKAM^(0)} provided that the Hessian of the
integrable part $h$ is non-degenerate on its open domain, say
$\Ascr\subseteq\reali^n$. Indeed, it is just matter of performing a
canonical change of coordinates that translates the origin of the
actions in correspondence to $\vet{I}^{\star}\in\Ascr$, because
$$
\frac{\partial h\big(\vet{I}\big)}{\partial I_j}\bigg|_{\vet{I}=\vet{I}^{\star}}=
\frac{\partial h\big(\vet{I}(\vet{p})\big)}{\partial p_j}\bigg|_{\vet{p}=\vet{0}}=
\omega^{(0)}_j
\quad\ \forall\ j=1,\ldots,n\,,
$$ where $\vet{I}=\vet{p}+\vet{I}^{\star}$. Obviously, the so called
action--frequency map in the integrable approximation, i.e.,
$\vet{I}^{\star}\mapsto\vet{\omega}^{(0)}$, can be inverted because
the Hessian of $h$ is non-degenerate. Therefore, the angular velocity
vector $\vet{\omega}^{(0)}$ can be used instead of $\vet{I}^\star$ in
order to parameterize the whole Hamiltonian.  Moreover the Fourier
decay of the coefficients with respect to the angles
$\vet{q}=\vet{\theta}$ allows to perform the
expansion~\eqref{frm:espansione-HKAM^(0)} in such a way that
$f_{\ell}^{(0,\,s)}=\Oscr(\epsilon^s)$. In other words, the positive
integer parameter $K$ can be chosen in such a way that the superscript
$s$ refers at the same time to both the order of magnitude and the
trigonometric degree (being $f_{\ell}^{(0,\,s)}\in\Pscr_{\ell,sK}$);
more details about that can be found in~\cite{Gio-Loc-1997}.

We are now ready for the description of the (generic) $r$-th step of
the normalization procedure, which defines the Hamiltonian $H^{(r)}$
starting from $H^{(r-1)}$, whose expansion is written as follows:
\begin{equation}
\label{frm:espansione-HKAM^(r-1)}
\vcenter{\openup1\jot 
\halign{
$\displaystyle\hfil#$&$\displaystyle{}#\hfil$&$\displaystyle#\hfil$\cr
  H^{(r-1)}(\vet p, \vet q) & \,=\,
  E^{(r-1)}+\scalprod {\vet \omega^{(r-1)}}{\vet p }+ 
  \sum_{s\ge 0}\sum_{\ell\ge 2} f_{\ell}^{(r-1,\,s)}(\vet p, \vet q)
  \cr
  & \,+\, \sum_{s\ge r}\sum_{\ell =0}^1 f_{\ell}^{(r-1,\,s)}(\vet p, \vet q)
  \ .
  \cr
}}
\end{equation}
Hereafter, we omit the dependence of the function from the parameters,
unless it has some special meaning. Let us assume that some
fundamental properties that hold true for $H^{(0)}$ are satisfied also
for the expansion above of $H^{(r-1)}$, i.e.,
$f_{\ell}^{(r-1,\,s)}\in\Pscr_{\ell,sK}\,$ and
$f_{\ell}^{(r-1,\,s)}=\Oscr(\epsilon^s)$. Since the $r$-th
normalization step aims to remove the main perturbing terms, that are
$f_{0}^{(r-1,\,r)}$ and $f_{1}^{(r-1,\,r)}$, we introduce a first
generating function $\chi_{1}^{(r)}$, that is determined by solving
the following (first) homological equation:
\begin{equation}
\poisson{\scalprod{ \vet \omega^{(r-1)}}{ \vet p}}{\chi^{(r)}_{1}}
+ f_{0}^{(r-1,\,r)}(\vet q) = \langle f_{0}^{(r-1,r)}(\vet q) \rangle_{\vet q}\ .
\label{frm:chi1rKAM}
\end{equation}
Since $f_{0}^{(r-1,\,r)}\in\Pscr_{0,rK}\,$, its expansion is written as
$$
f_{0}^{(r-1,\,r)}( \vet q)=\sum_{{\scriptstyle{| \vet k|\le rK}}}
c_{\vet k} \exp\big(\imunit\scalprod{\vet k}{ \vet q}\big)\ ,
$$
where the complex coefficients are such that
$c_{-\vet k}={\bar c}_{\vet k}\,$. Therefore, one can easily
check that the first homological equation~\eqref{frm:chi1rKAM}
is solved by putting 
$\langle f_{0}^{(r-1,r)}(\vet q) \rangle_{\vet q}=c_{\vet 0}$ and
\begin{equation}
\chi^{(r)}_{1}( \vet q)=\sum_{{\scriptstyle{0<| \vet k|\le rK}}}
  \frac{c_{\vet k} \exp\big(\imunit\scalprod{\vet k}{ \vet q}\big)}
  {\imunit\scalprod{\vet k}{\vet\omega^{(r-1)}}}\ .
\label{frm:espansione-chi1rKAM}
\end{equation}
In order to preserve the validity of the solution above, of course,
we have to require that none of the divisors can eventually vanish;
thus we assume the following non-resonance condition:
\begin{equation}
  \scalprod{\vet k}{\vet\omega^{(r-1)}} \neq 0
  \quad\ \forall\ 0<\vet{k}\le rK\ .
\label{frm:nonres-cond-passor-KAM-primomezzopasso}
\end{equation}
The first half of the $r$-th normalization step is completed by
introducing $\hat H^{(r)}=\exp\big(\Lie{\chi_1^{(r)}}\big)H^{(r-1)}$.
Such an intermediate Hamiltonian can be written in a form similar to
formula~\eqref{frm:espansione-HKAM^(r-1)}, i.e.,
\begin{equation}
\label{frm:espansione-hatHKAM^(r)}
\vcenter{\openup1\jot 
\halign{
$\displaystyle\hfil#$&$\displaystyle{}#\hfil$&$\displaystyle#\hfil$\cr
  \hat H^{(r)}(\vet p, \vet q) =
  & E^{(r)}+ \scalprod {\vet \omega^{(r)}}{\vet p }+ 
  \sum_{s\ge 0}\sum_{\ell\ge 2} \hat f_{\ell}^{(r,\,s)}(\vet p, \vet q)+
  \sum_{s\ge r}\sum_{\ell =0}^1 \hat f_{\ell}^{(r,\,s)}(\vet p, \vet q)
  \ ,
  \cr
}}
\end{equation}
where the recursive definitions of the new summands $\hat
f_{\ell}^{(r,\,s)}$ (in terms of $f_{\ell}^{(r-1,\,s)}$) can be given
by exploiting the linearity of the Lie series and by separating the
functions according to the different classes $\Pscr_{\ell,sK}$ they
belong to. We think it is convenient to formulate these definitions in
a rather unconventional way, by using a notation similar to that
commonly used in the programming languages; in our opinion, such a
choice should make easier the translation of the formal algorithm in
any code to be executed in a computational environment. For this
purpose, we first define\footnote{We remark that
  $f_{\ell}^{(r-1,\,s)}$ do not enter in the
  expansion~\eqref{frm:espansione-HKAM^(r-1)} if $\ell=0,\,1$ and
  $s<r$. The same applies to the terms $\hat f_{\ell}^{(r,\,s)}$ that
  do not make part of the expression of $\hat H^{(r)}$, which is
  written in~\eqref{frm:espansione-hatHKAM^(r)}. However, the
  recursive definitions described in the present subsection are such
  that $f_{\ell}^{(r-1,\,s)}=\hat f_{\ell}^{(r,\,s)}=0$ $\forall\ 0\le
  s<r$ when $\ell=0,\,1$.} $\hat f_{\ell}^{(r,\,s)}(\vet p, \vet q)=
f_{\ell}^{(r-1,\,s)}(\vet p, \vet q)$ $\forall\ \ell\ge 0,\,s\ge 0$.
Then, \emph{by abuse of notation}, we update $\lfloor s/r\rfloor$
times the definition of the terms $\hat f_{\ell}^{(r,\,s)}$ appearing
in the expansion of the new Hamiltonian according to the following
rule:
\begin{equation}
  \label{frm:ridefinizioni-hatHKAM^(r)}
  \hat f_{\ell -j}^{(r,\,s+jr)} \hookleftarrow
  \frac{1}{j!} \lie{\chi_1^{(r)}}^j f_{\ell}^{(r-1,\,s)}
  \quad \forall \  \ell \ge 1, \ 1\le j \le \ell,
  \ s \ge 0\ ,
\end{equation}
where with the notation $a \hookleftarrow b$ we mean that the quantity
$a$ is redefined so as to be equal to $a+b$. Moreover, there is a last
additional contribution that is due to the application of the Lie
series to the Hamiltonian $H^{(r-1)}$, and in order to take it into
account we write
\begin{equation}
  \label{frm:ridefinizione-utile-x-prima-eq-omolKAM}
  \hat f_{0}^{(r,\,r)} \hookleftarrow
  \lie{\chi_1^{(r)}}\scalprod {\vet \omega^{(r-1)}}{\vet p }\ .
\end{equation}
However, because of the homological equation~\eqref{frm:chi1rKAM}, we
can finally put $\hat f_{0}^{(r,\,r)}=0$ and update the constant
energy value so that
\begin{equation}
  \label{frm:ridefinizione-utile-x-energia}
  E^{(r)}=E^{(r-1)}+
  \langle f_{0}^{(r-1,\,r)}\rangle_{\vet q}\ .
\end{equation}
At this point, it is important to remark that the angular average of
the remaining perturbing term that is $\Oscr(\epsilon^r)$, i.e.,
$\langle\hat f_{1}^{(r,\,r)}\rangle_{\vet q}$ is exactly of the same
type as $\scalprod{\vet \omega^{(r-1)}}{\vet p }$ (this means that
both of them are linear with respect to the actions and do not depend
on the angles). Therefore, it is useful to update also the angular
velocity vector\footnote{We emphasize that this is one of the main
  differences with respect to the original proof scheme designed by
  Kolmogorov, where the angular velocity vector is kept fixed at every
  normalization step (see~\cite{Ben-Gal-Gio-Str-1984} for a fully
  consistent translation of such an approach, that is implemented by
  using the Lie series technique).} by joining together these two
terms.  This can be done, by redifining
\begin{equation}
  \label{frm:ridefinizione-utile-x-vel-ang}
  \scalprod {\vet \omega^{(r)}}{\vet p } =
  \scalprod {\vet \omega^{(r-1)}}{\vet p } +
  \langle \hat f_{1}^{(r,\,r)}\rangle_{\vet q}
\end{equation}
and
\begin{equation}
  \label{frm:ridefinizione-utile-x-seconda-eq-omolKAM}
  \hat f_{1}^{(r,\,r)} =  \hat f_{1}^{(r,\,r)} -
  \langle \hat f_{1}^{(r,\,r)}\rangle_{\vet q}\ .
\end{equation}

Let us recall that all the terms $\hat f_{\ell}^{(r,\,s)}$ that appear
in formula~\eqref{frm:espansione-hatHKAM^(r)} are organized so that
they belong to different classes of functions. In order to prove that
these structures are suitably preserved by the normalization algorithm,
the following statement is essential.

\begin{lemma}
  \label{lem:classi-funzioni}
  Let us consider two generic functions $g\in\Pscr_{\ell,sK}$ and
  $h\in\Pscr_{m,rK}\,$, where $K$ is a fixed positive integer
  number. Then, the following inclusion property holds
  true\footnote{The statement can be considered as valid also in the
    trivial case with $\ell=m=0$, by enlarging the definition of the
    classes of functions so that
    $\Pscr_{-1,sK}=\big\{0\big\}\ \forall\>s\in\naturali$.}:
  $$
  \big\{g,h\big\} = \Lie{h}\,g \in \Pscr_{\ell+m-1,(r+s)K}
  \quad
  \ \forall\>\ell,\,m,\,r,\,s\in\naturali\ .
  $$
\end{lemma}
The proof is omitted, because it can be obtained as a straightforward
consequence of the definition of the Poisson brackets. By applying
repeatedly the lemma above and a trivial induction argument to
formul{\ae}~\eqref{frm:ridefinizioni-hatHKAM^(r)}--\eqref{frm:ridefinizione-utile-x-seconda-eq-omolKAM},
one can easily prove that $E^{(r)}\in\Pscr_{0,0}$ and $\hat
f_{\ell}^{(r,\,s)}\in\Pscr_{\ell,sK}$ for all the terms of type $\hat
f_{\ell}^{(r,\,s)}$ that appear in
formula~\eqref{frm:espansione-hatHKAM^(r)}. Moreover, it can be
ensured that $\big|E^{(r)}-E^{(r-1)}\big|=\Oscr(\epsilon^r)$
and $\hat f_{\ell}^{(r,\,s)}=\Oscr(\epsilon^s)$, if the same relation
is assumed to be true at the end of the previous normalization step,
i.e., $f_{\ell}^{(r-1,\,s)}=\Oscr(\epsilon^s)$.

In order to complete the $r$-th normalization step, we have to remove
the remaining perturbing term that is $\Oscr(\epsilon^r)$ and appears
in the expansion~\eqref{frm:espansione-hatHKAM^(r)} of Hamiltonian
$\hat H^{(r)}$, i.e., $\hat f_{1}^{(r,\,r)}$. For such a purpose, we
determine a second generating function $\chi_{2}^{(r)}$, by solving
the following (second) homological equation:
\begin{equation}
\poisson{\scalprod{ \vet \omega^{(r)}}{ \vet p}}{\chi^{(r)}_{2}}
+ \hat f_{1}^{(r,\,r)}(\vet p,\vet q) = 0\ .
\label{frm:chi2rKAM}
\end{equation}
We can deal with the equation above in a very similar way with respect
to what has been done for the first homological
equation~\eqref{frm:chi1rKAM}. In fact, the solution
of~\eqref{frm:chi2rKAM} can be written as follows:
\begin{equation}
  \chi^{(r)}_{2}(\vet p,\vet q)=
  \sum_{{\scriptstyle{|\vet j|=1}}}\,
  \sum_{{\scriptstyle{0<| \vet k|\le rK}}}
  \frac{c_{\vet j,\vet k}\, {\vet p}^{\vet j}
    \exp\big(\imunit\scalprod{\vet k}{ \vet q}\big)}
  {\imunit\scalprod{\vet k}{\vet\omega^{(r)}}}\ ,
\label{frm:espansione-chi2rKAM}
\end{equation}
where the expansion of the perturbing term
$\hat f_{1}^{(r,\,r)}\in\Pscr_{1,rK}$ is of type 
$$
\hat f_{1}^{(r,\,r)}(\vet p,\vet q)=
  \sum_{{\scriptstyle{|\vet j|=1}}}\,
  \sum_{{\scriptstyle{0<| \vet k|\le rK}}}
  c_{\vet j,\vet k}\, {\vet p}^{\vet j}
    \exp\big(\imunit\scalprod{\vet k}{ \vet q}\big)\ .
$$
Let us recall that the angular average of $\hat f_{1}^{(r,\,r)}$ is
equal to zero, because of the
redefinition~\eqref{frm:ridefinizione-utile-x-seconda-eq-omolKAM}.  Of
course, the solution written in~\eqref{frm:espansione-chi2rKAM} is
valid provided that the following non-resonance condition is
satisfied:
\begin{equation}
  \scalprod{\vet k}{\vet\omega^{(r)}} \neq 0
  \quad\ \forall\ 0<\vet{k}\le rK\ .
\label{frm:nonres-cond-passor-KAM-secondomezzopasso}
\end{equation}
Finally, $H^{(r)}=\exp\big(\Lie{\chi_2^{(r)}}\big)\hat H^{(r)}$ is the
new Hamiltonian that is defined by the canonical transformation of
coordinates that is introduced by the $r$-th normalization step.  Also
the expansion of such a Hamiltonian can be written in a form similar
to~\eqref{frm:espansione-HKAM^(r-1)}, i.e.,
\begin{equation}
\label{frm:espansione-HKAM^(r)}
\vcenter{\openup1\jot 
\halign{
$\displaystyle\hfil#$&$\displaystyle{}#\hfil$&$\displaystyle#\hfil$\cr
  H^{(r)}(\vet p, \vet q) =
  & E^{(r)}+ \scalprod {\vet \omega^{(r)}}{\vet p }+ 
  \sum_{s\ge 0}\sum_{\ell\ge 2} f_{\ell}^{(r,\,s)}(\vet p, \vet q)+
  \sum_{s\ge r+1}\sum_{\ell =0}^1 f_{\ell}^{(r,\,s)}(\vet p, \vet q)
  \ .
  \cr
}}
\end{equation}
In this case too, the recursive definitions of the new summands
$f_{\ell}^{(r,\,s)}$ can be given by exploiting the linearity of the
Lie series and by separating the functions according to the different
classes they belong to. Let us start by introducing
$f_{\ell}^{(r,\,s)}(\vet p, \vet q)=\hat f_{\ell}^{(r,\,s)}(\vet p,
\vet q)$ $\forall\ \ell\ge 0,\,s\ge 0$.  By a new abuse of notation,
we update many times the definition of the terms appearing in the
expansion of Hamiltonian $H^{(r)}$ according to the following rule:
\begin{equation}
  \label{frm:ridefinizioni-HKAM^(r)}
  f_{\ell}^{(r,\,s+jr)} \hookleftarrow
  \frac{1}{j!} \lie{\chi_2^{(r)}}^j \hat f_{\ell}^{(r,\,s)}
  \quad \forall\ \ell \ge 2, \ j \ge 1,\ s \ge 0
  \ {\rm or}\ \forall\ \ell =0,1, \ j \ge 1,\ s > r
  \ .
\end{equation}
In order to take into account also the summands that are generated by
the application of the Lie series $\exp\big(\Lie{\chi_2^{(r)}}\big)$
to both the terms $\scalprod {\vet \omega^{(r)}}{\vet p }$ and
$f_{1}^{(r,\,r)}(\vet p, \vet q)$, we add the prescription
\begin{equation}
  \label{frm:ridefinizione-utile-dovuta-a-seconda-eq-omolKAM}
  f_{1}^{(r,\,(j+1)r)} \hookleftarrow
  \frac{j}{(j+1)!} \lie{\chi_2^{(r)}}^j \hat f_{\ell}^{(r,\,r)}
  \quad \forall\ j \ge 1\ ,
\end{equation}
where we make use of formula~\eqref{frm:chi2rKAM}. Also the last
redefinition, i.e.,
\begin{equation}
  f_{1}^{(r,r)} = 0\ ,
  \label{frm:azzeramento-dovuto-a-seconda-eq-omolKAM}
\end{equation}
is a straightforward consequence of the second homological equation.
By applying again Lemma~\ref{lem:classi-funzioni} and a trivial
induction argument to
formul{\ae}~\eqref{frm:ridefinizioni-HKAM^(r)}--\eqref{frm:ridefinizione-utile-dovuta-a-seconda-eq-omolKAM},
one can easily prove that $f_{\ell}^{(r,\,s)}\in\Pscr_{\ell,sK}$ for
all the summands $f_{\ell}^{(r,\,s)}=\Oscr(\epsilon^s)$ that appear in
formula~\eqref{frm:espansione-HKAM^(r)}.

This final remark ends the description of the $r$-th normalization
step of the algorithm that can be iterated so as to determine the next
Hamiltonian $H^{(r+1)}$, starting from $H^{(r)}$, and so on.

Let us add a few further comments about the algorithm constructing the Kolmogorov
normal form in order to understand its applicability. In practice,
one is often interested in determining an approximation up to a fixed
order, say $R_{\rm I}\in\naturali$, of the motions travelling an
invariant KAM torus. For this purpose, starting from $H^{(0)}$, one
has to preliminarly compute the Taylor-Fourier truncated expansions of
the following type, for all the Hamiltonian $H^{(r)}$ that are
introduced by the normalization algorithm with
$r=1,\,\ldots\,,\,R_{\rm I}\,$:
\begin{equation}
\label{frm:espansione-appr-HKAM^(r)}
  H^{(r)}(\vet p, \vet q) \simeq
  E^{(r)}+ \scalprod {\vet \omega^{(r)}}{\vet p }+ 
  \sum_{s=0}^{R_{\rm I}}\sum_{\ell= 0}^{\ell_{\rm max}} f_{\ell}^{(r,\,s)}(\vet p, \vet q)
  \ ,
\end{equation}
where all the terms that are $o\big(\epsilon^{R_{\rm I}}\big)$ or of
polynomial degree larger than $\ell_{\rm max}$ with respect to the
actions\footnote{In the practical applications, it is very common to
  truncate this kind of Taylor series expansions up to a finite
  degree. In this framework, it is important to remark that the upper
  limit on the degree in actions is preserved by the Lie series having
  $\chi_1^{(r)}\in\Pscr_{0,rK}$ and $\chi_2^{(r)}\in\Pscr_{1,rK}$ as
  generating functions. This can be easily checked by applying
  repeatedly Lemma~\ref{lem:classi-funzioni}, that can be used also to
  prove that just functions of type $f_{\ell}^{(r,\,s)}$ with $\ell\le
  R_{\rm I}+1$ are involved in the definitions of $\chi_1^{(r)}$ and
  $\chi_2^{(r)}$ $\forall\ r=1,\,\ldots\,,\,R_{\rm I}\,$. In other
  terms, this means that the request of determining an approximation
  up to a fixed order of magnitude $\Oscr\big(\epsilon^{R_{\rm
      I}}\big)$ (for what concerns the canonical transformation that
  conjugates some orbits to an invariant torus) yields in a fully
  consistent way also a truncation limit on the polynomial degree in
  the actions.} have been neglected. Let us recall that the algorithm
works in such a way to define $f_{\ell}^{(r,\,s)}=0$
$\forall\ \ell=0,1\,,\ 0\le s\le r$. When the first $R_{\rm I}$
normalization steps are performed, all the generating functions
$\chi_1^{(r)}$ and $\chi_2^{(r)}$ $\forall\ r=1,\,\ldots\,,\,R_{\rm
  I}\,$ are fully determined. Their composition allows to compute the
expansion of $\psi_\epsilon$ that enters in the definition of the
semi-analytic scheme of integration~\eqref{frm:semi-anal-scheme} and
is truncated, once again, so as to neglect all the summands that are
$o\big(\epsilon^{R_{\rm I}}\big)$. Therefore, the wanted approximation
of the motions travelling an invariant KAM torus up to a fixed order
of magnitude $\Oscr\big(\epsilon^{R_{\rm I}}\big)$ can be provided by
the scheme~\eqref{frm:semi-anal-scheme} where also the normal form
Hamiltonian $\Kscr$ is replaced by $H^{(R_{\rm I})}$, which requires
$\ell_{\rm max}\big(R_{\rm I}+1)^2$ functions of type
$f_{\ell}^{(r,\,s)}\in\Pscr_{\ell,sK}$ to be determined. Since their
expansions in Taylor-Fourier series are finite (recall
definition~\eqref{frm:esempio-g-in-Pscr}), all their coefficients are
\emph{representable on a computer} (that is equipped with a large
enough memory). Therefore, it is \emph{finite} also the number of
elementary operations that are defined by the Poisson brackets
prescribed by normalization algorithm. The same conclusion applies
also for the aforementioned expansion of the canonical transformation
$\psi_\epsilon\,$. As a whole, we can conclude that the wanted
approximation of the motions travelling an invariant KAM torus is
\emph{explicitly computable}, because the total amount of operations
that are defined by the normalization algorithm is \emph{finite}.

\subsection{On the convergence of the algorithm constructing the Kolmogorov normal form}\label{sbs:conv-Kolmog-original}
In the present context, it is useful to introduce another
version of the KAM theorem.
\begin{proposition}
\label{teorema-KAM-versione-Poschel-Chiara}
Consider the family of Hamiltonians $H^{(0)}(\vet p, \vet
q;\vet\omega^{(0)})$ of the type described
in~\eqref{frm:espansione-HKAM^(0)}. Those functions are defined so
that $H^{(0)}:\,\Ascr\times\toro^{n}\times\Uscr\mapsto\reali$, where
both $\Ascr$ and $\Uscr$ are open subsets of $\reali^n$, being
$\vet{0}\in\Ascr$ and $\Uscr$ bounded. Therefore, $(\vet p, \vet q)$
are action-angle canonical coordinates and the family of Hamiltonians
is parameterized with respect to $\vet\omega^{(0)}\in\Uscr$.  Let us
also assume that for some fixed and positive values of
$K\in\naturali$, $\epsilon\in\reali$ and $E\in\reali$, the following
inequalities are satisfied by the functions
$f_\ell^{(0,s)}\in\Pscr_{\ell,sK}\,$:
\begin{equation}
  \sup_{(\vet p,\vet q;\vet\omega^{(0)})\in
    \Ascr\times\toro^{n}\times\Uscr}
  \left|f_\ell^{(0,s)}(\vet p,\vet q;\vet\omega^{(0)})\right|
  \le E\,\epsilon^s
 \end{equation}
$\forall\ s\ge 1,\ \ell\ge 0$ and $\forall\ \ell\ge 2\ {\rm when}\ s=0$.

\noindent
Then, there is a positive $\epsilon^{\star}$ such that for $0\le
\epsilon<\epsilon^{\star}$ the following statement holds true: there
exists a non-resonant set $\Uscr^{(\infty)}\subset\Uscr$ such that the
Lebesgue measure $\mu$ of the complementary set
$\Uscr\setminus\Uscr^{(\infty)}$ goes to zero for $\epsilon\to 0$ and
for each $\vet \omega^{(0)}\in\Uscr^{(\infty)}$ there is an analytic
canonical transformation $(\vet p,\vet
q)=\psi_{\epsilon;\vet\omega^{(0)}}^{(\infty)}(\vet P,\vet Q)$ leading
the Hamiltonian to the normal form
\begin{equation}
\vcenter{\openup1\jot\halign{
 \hbox {\hfil $\displaystyle {#}$}
&\hbox {\hfil $\displaystyle {#}$\hfil}
&\hbox {$\displaystyle {#}$\hfil}\cr
H^{(\infty)}(\vet P,\vet Q;\vet\omega^{(0)}) &=
&E^{(\infty)} + \vet \omega^{(\infty)}\cdot \vet P
+\sum_{s\ge 0}\sum_{\ell\ge 2}f_{\ell}^{(\infty,\,s)}(\vet P, \vet Q;\vet\omega^{(0)})
\ ,
\cr
}}
\end{equation}
where $f_{\ell}^{(\infty,\,s)}\in\Pscr_{\ell,sK}$ $\forall\ s\ge
0,\ \ell\ge 2$ and $E^{(\infty)}$ is a finite real value fixing
the constant energy level that corresponds to the invariant torus
$\big\{( \vet P = \vet 0,\, \vet Q\in\toro^{n})\big\}\,$.  Moreover,
the canonical change of coordinates is close to the identity in the
sense that $\big\|\psi_{\epsilon;\vet\omega^{(0)}}^{(\infty)}(\vet P,
\vet Q)-(\vet P, \vet Q)\big\|=\Oscr(\epsilon)$ and the same applies
also to both the energy level and the detuning of the angular velocity
vector (that are
$\big|E^{(\infty)}-E^{(0)}\big|=\Oscr(\epsilon)$ and
$\big\|\omega^{(\infty)}-\omega^{(0)}\big\|=\Oscr(\epsilon)$,
respectively).
\end{proposition}
The statement above is substantially equivalent to that claimed in
theorem~C of~\cite{Poschel-1989} (which is considered as a classical
version of the KAM theorem, in the very own words of the Author,
J. P{\"o}schel). The proof of
Proposition~\ref{teorema-KAM-versione-Poschel-Chiara} can be obtained
by adapting the one described in~\cite{Caracciolo-2021} in such a way
to prove the convergence of the normalization algorithm described in
the previous Subsection~\ref{sbs:Kolmog-original}. Indeed, both
articles~\cite{Poschel-1989} and~\cite{Caracciolo-2021} deal only with
the more complicate proof of existence for invariant tori that are of
dimension smaller than the number $n$ of degrees of freedom and have
elliptic character in the transverse directions. The construction of
the normal form corresponding to such a type of invariant manifolds
will be widely discussed in the next Section~\ref{sec:kolmog}. As a
main difference between the approaches developed in those two works,
let us recall that the proof adopted in~\cite{Poschel-1989} is based
on a fast convergence scheme of quadratic type (a so called
Newton-like method, where perturbing terms of order of magnitude
$\Oscr\big(\epsilon^{2^{r-1}}\big)$ are removed during the $r$-th
normalization step). Such a technique has been adopted since the very
first works in KAM theory, but the convergence of the normalization
algorithm described in Subsection~\ref{sbs:Kolmog-original} is of
linear type (because perturbing terms of order of magnitude
$\Oscr\big(\epsilon^{r}\big)$ are removed during the $r$-th
normalization step). The latter is in a better position for the
applications\footnote{This is the main reason why the present work is
  focusing on approaches based on a convergence scheme of linear
  type. A very far from being exhaustive list of references to
  applications of KAM theorem has been discussed in the Introduction.}
and a complete proof of the KAM theorem adopting a convergence method
of linear type is available since the last decade of the past century
(see~\cite{Gio-Loc-1997}). Rather curiously, the best way to translate
the algorithm constructing the Kolmogorov normal form in a
computer-assisted proof requires to join the convergence scheme of
linear type (in order to explicitly perform on a computer the largest
possible number $R_{\rm I}$ of preliminary steps) with that of
quadratic type (that provides a statement of KAM theorem that is very
suitable to rigorously complete the proof).  This is one of the main
conclusions discussed in a recent work (see~\cite{Val-Loc-2021}).

The statement of Proposition~\ref{teorema-KAM-versione-Poschel-Chiara}
highlights that we are forced to provide a result which holds true with
respect to the Lebesgue measure, because we have \emph{chosen} to
adopt a version of the normalization algorithm where the angular
velocity vector is allowed to vary at each step (recall
formula~\eqref{frm:ridefinizione-utile-x-vel-ang} that defines the
detuning shift $\vet \omega^{(r)}-\vet \omega^{(r-1)}$). This means
that such a statement has to
be understood in a probabilistic sense, because we are not able to describe in detail the structure of
the non-resonant set $\Uscr^{(\infty)}$. In particular, for a fixed
initial value of the angular velocity vector $\vet \omega^{(0)}$ we
cannot establish whether the specific Hamiltonian $H^{(0)}(\vet p,
\vet q;\vet\omega^{(0)})$ can be brought in Kolmogorov normal form or
not. We can just claim that the normalization algorithm can converge
with a rate of success (i.e.,
$\mu\big(\Uscr\setminus\Uscr^{(\infty)}\big)\,/\,\mu(\Uscr)$) that
gets larger and larger when the small parameter $\epsilon$ which rules
the size of the perturbation is decreasing. On the other hand, we can
characterize very well the set of the final values of the angular
velocities, i.e.,
$\big\{\vet\omega^{(\infty)}\big(\vet\omega^{(0)}\big)\,:\ \vet\omega^{(0)}\in\Uscr^{(\infty)}\big\}$,
because they are Diophantine. In the recent work~\cite{San-Dan-2021},
the problem of the convergence of this type of normalization
algorithms is revisited so as to provide a KAM-like statement. It is
proved by fixing since the beginning the final value
$\vet\omega^{(\infty)}$ and its non-resonance properties (that allow
to explicitly solve the homological equations at every step of the
algorithm). Moreover, the total detuning
$\vet\omega^{(\infty)}-\vet\omega^{(0)}$ is given in terms of series whose
coefficients are defined in a recursive way. Therefore, the
convergence of the normalization algorithm is ensured (provided that
the perturbation is small enough), the total detuning is estimated
explicitly, while the exact location of $\vet\omega^{(0)}$ remains
partially unknown, because it can be determined just by iterating
\emph{ad infinitum} the computational procedure.

\section{Construction of invariant elliptic tori by a normal form algorithm}\label{sec:kolmog}

Elliptic tori are compact invariant manifolds of dimension smaller
than the maximal one, that is equal to the number $n$ of degrees of
freedom.  In order to better imagine them, let us consider a phase
space $\Fscr$ that is endowed by the canonical coordinates $( \vet P,
\vet Q, \vet X, \vet Y)$, where $( \vet P, \vet Q)\in
\reali^{n_1}\times \toro^{n_1}$ are action-angle variables and also
$(\vet X, \vet Y)\in \reali^{n_2} \times \reali^{n_2}$ denote pairs of
conjugate (momenta and) coordinates, while $n=n_1+n_2$ with both $n_1$
and $n_2$ positive integers. Let us consider a Hamiltonian of the
following type:
$$
\Hscr( \vet P, \vet Q, \vet X, \vet Y) = \scalprod {\vet \omega}{\vet P}+ 
\sum_{j=1}^{n_2} \frac{\Omega_j}{2} (X_j^2 + Y_j^2)+
\Rscr( \vet P, \vet Q, \vet X, \vet Y)\ ,
$$
where $\vet\Omega \in \reali^{n_2}$ and the remainder $\Rscr$ is an analytic function with respect to
its arguments and is such that
$\Rscr( \vet P, \vet Q, \vet X, \vet Y)=o\big(\|\vet P\|+
\|(\vet X,\vet Y)\|^2\big)$, when $( \vet P, \vet X, \vet Y)\to( \vet
0, \vet 0, \vet 0)$. It is easy to check that
\begin{equation}
  \label{frm:soluzione-su-toro-ellittico}
  ( \vet  P(t),  \vet  Q(t),  \vet X(t),  \vet Y(t)) =
  \big(  \vet 0,  \vet  Q(0) +  \vet  \omega t,  \vet  0,  \vet 0 \big)
\end{equation}
is a solution of Hamilton equations, since the function $\Hscr$,
except for its main part, contains terms of type $\Oscr(\| \vet P
\|^2)$, $\Oscr(\|\vet P\|\|(\vet X, \vet Y)\|)$ and $\Oscr(\|(\vet
X,\vet Y)\|^3)$ only.  Because of this remark, it is evident that the
$n_1$--dimensional manifold $\big\{( \vet P, \vet Q, \vet X, \vet
Y)\,:\ \vet P =\vet 0,\>\vet Q \in\toro^{n_1},\>\vet X =\vet Y =\vet 0
\big\}$ is invariant. The elliptical character is given by the fact
that, in the remaining $n_2=n-n_1$ degrees of freedom, the dynamics
that is transverse with respect to such an invariant manifold is given
by the composition of $n_2$ oscillatory motions whose periods tend to
the values $2\pi/\Omega_1\,,\,\ldots\,,\,2\pi/\Omega_{n_2}\,$, in the
limit of $( \vet P, \vet X, \vet Y)\to( \vet 0, \vet 0, \vet 0)$.  Of
course, this is due to the occurrence of the term
$\sum_{j=1}^{n_2}\Omega_j(X_j^2 + Y_j^2)/2$ which overwhelms the
effect of the remainder $\Rscr$ in the so called limit of small
oscillations.

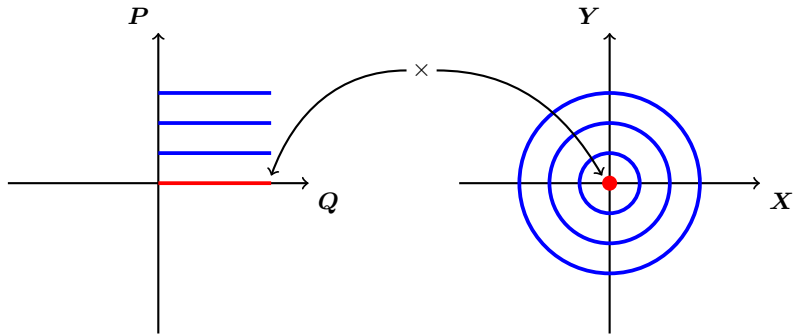
\begin{figure}
  \begin{center}
  \begin{tikzpicture}
    \draw[thick,->] (-2,0) -- (2,0) node[anchor=north west] {$\vet Q$};
    \draw[thick,->] (0,-2) -- (0,2) node[anchor=south east] {$\vet P$};

    \draw[red,line width=0.5mm] (0,0)--(1.5,0);
    \draw[blue,line width=0.5mm] (0,0.4)--(1.5,0.4);
    \draw[blue,line width=0.5mm] (0,0.8)--(1.5,0.8);
    \draw[blue,line width=0.5mm] (0,1.2)--(1.5,01.2);
    
    \draw[thick,->] (4,0) -- (8,0) node[anchor=north west] {$\vet X$};
    \draw[thick,->] (6,-2) -- (6,2) node[anchor=south east] {$\vet Y$};

    \coordinate (O) at (6,0);
    \fill[red] (6,0)  circle[radius=1mm];
    \draw[blue,line width=0.5mm] (O) circle (0.4);
    \draw[blue,line width=0.5mm] (O) circle (0.8);
    \draw[blue,line width=0.5mm] (O) circle (1.2);

    \coordinate (X) at (3.5,1.5);
    \node at  (X) {$\times$};
    \draw[thick,->]    (3.3,1.5) to[out=180,in=70] (1.5,0.1);
    \draw[thick,->]    (3.7,1.5) to[out=0,in=120] (5.9,0.1);
  \end{tikzpicture}
  \end{center}

  \caption{Schematic representation of an elliptic torus.  The orbit
    is given by the cartesian product of the two invariant surfaces
    that are marked in red, that are a torus (in the left panel) and a
    stable equilibrium point (on the right, resp.).}
  \label{fig:elltor-scheme}
\end{figure}

The name of elliptic torus is well justified by all the remarks
discussed since the beginning of the present section. A schematic
representation of such kind of invariant manifolds is sketched
in Figure~\ref{fig:elltor-scheme}.

\subsection{Algorithmic construction of the normal form for elliptic tori}\label{sbs:Kolmog-4-ell-tori}

Since we aim at introducing the algorithm constructing the normal form
for invariant elliptic tori in a way that is as much as possible
coherent with what we have already done in
Subsection~\ref{sbs:Kolmog-original} for KAM tori, we prefer to not
adopt canonical coordinates $( \vet p, \vet q, \vet x, \vet y)\in
\reali^{n_1}\times\toro^{n_1}\times\reali^{n_2}\times\reali^{n_2}$
that are substantially the ones considered in the discussion at the
beginning of the present section. Indeed, we think it is convenient to
introduce the so called action-angle coordinates for harmonic
oscillators, in order to replace the polynomial ones, that are $(\vet
x, \vet y)\in\reali^{n_2}\times\reali^{n_2}$; this means that we
define $(\vet J, \vet\phi)\in\big(\reali_+^{n_2}\cup\{\vet
0\}\big)\times\toro^{n_2}$ so that $x_j=\sqrt{2J_j}\cos\phi_j$ and
$y_j=\sqrt{2J_j}\sin\phi_j\,$, where this change of coordinates is
canonical $\forall\ j=1,\,\ldots\,,\,n_2\,$.

We are now ready to introduce classes of functions depending on $(
\vet p, \vet q, \vet J, \vet\phi)\in
\reali^{n_1}\times\toro^{n_1}\times\big(\reali_+^{n_2}\cup\{\vet
0\}\big)\times\toro^{n_2}$ in a very similar way to what has been
previously done.
For some fixed positive integer $K$
we introduce the distinct classes of functions $\PPset_{\hat
  m,\,\hat \ell,\,sK}\,$, with integers $\hat m,\,\hat \ell,\,s\ge
0\,$; any generic function $g\in\PPset_{\hat m,\,\hat \ell,\,sK}$
can be written as
\begin{equation}
\vcenter{\openup1\jot\halign{
 \hbox {\hfil $\displaystyle {#}$}
 &\hbox {$\displaystyle {#}$\hfil}\cr
 g &( \vet p, \vet q, \vet J,\vet\phi) =
 \cr
 &\sum_{{\scriptstyle{\vet m\in\naturali^{n_1}}}\atop{\scriptstyle{|\vet m|=\hat m}}}
 \sum_{{\scriptstyle{\vet\ell\in\naturali^{n_2}}}\atop{\scriptstyle{| \vet \ell|=\hat{\ell}}}}
 \sum_{{\scriptstyle{{ \vet k\in\interi^{n_1}}}\atop{\scriptstyle{| \vet k|\le sK}}}} 
 \,\sum_{{\scriptstyle{\hat\ell_j=-\ell_j,\,-\ell_j+2,\ldots,\,\ell_j}}\atop{\scriptstyle{j=1,\,\ldots\,,n_2}}}
 c_{\vet m,\vet \ell,\vet k,\hat {\vet \ell}}\,
 \vet p^{\vet m} \big(\sqrt{\vet{J}}\big)^{\vet \ell}
 \exp\big[\imunit(\scalprod{\vet k}{ \vet q}
   +\scalprod{\hat{\vet\ell}}{\vet\phi})\big] \>,
 \cr
}}
\label{frm:esempio-g-in-PPset}
\end{equation}
where the complex coefficients are such that $c_{\vet m,\vet
  \ell,-\vet k,-\hat{\vet \ell}}={\bar c}_{\vet m,\vet \ell,\vet
  k,\hat {\vet \ell}}$, then the codomain of any $g\in\PPset_{\hat
  m,\,\hat \ell,\,sK}$ is included in $\reali$. Let us emphasize that,
in each term appearing in the Taylor-Fourier expansion of a function
belonging to a class of type $\PPset_{\hat m,\,\hat \ell,\,sK}$, the
indexes vector $(\hat{\ell}_1\,,\,\ldots\,,\,\hat{\ell}_{n_2})$ are
subject to special restrictions that are inherited by the
corresponding polynomial structure with respect to the variables
$(\vet x,\vet y)=\big(\sqrt{2\vet J}\cos\vet\phi,\sqrt{2\vet
  J}\sin\vet\phi\big)$. In fact, they are such that
$\forall\ j=1,\,\ldots\,,n_2$ the $j$-th component of the Fourier
harmonic $\hat{\ell}_j$ must have the same parity with respect to the
correponding degree $\ell_j$ of $\sqrt{J_j}$ and also the inequality
$\big|\hat{\ell}_j\big|\le\ell_j$ must be satisfied\footnote{When
  there are variables such that they appear in the Taylor-Fourier
  expansions of a function so that they follow this kind of
  restrictions, then they are often said to be of D'Alembert
  type. { This name is given by analogy, because in Celestial
    Mechanics the secular part of the Hamiltonian perturbing terms due
    to the interactions between planets shows the same kind of
    expansions, since they satisfy the so called D'Alembert
    rules.}}.  Furthermore, we will say that $g\in \Pset_{\ell,sK}$ if
\begin{equation}
  g \in
  \bigcup_{\hat m \ge 0, \hat \ell \ge 0 \atop{ 2\hat m + \hat \ell = \ell}}
  \PPset_{\hat m , \hat \ell,sK}\ .
\label{frm:def-PPset}
\end{equation}
In other words, a function belonging to the class $\Pset_{\ell,sK}$ depends
on the actions so as to be homogeneous polynomials of total degree
$\ell$ in the square roots of $\vet{p}$ and $\vet J$, while its
Fourier expansion contain harmonics of total trigonometric degree in
$\vet q$ that are not larger than $sK$.

In order to extend the approach described in
Subsection~\ref{sbs:Kolmog-original} with the aim to design an
efficient algorithm constructing the normal form in the case of
elliptic tori, we are also forced to reformulate the
Lemma~\ref{lem:classi-funzioni} in a suitable version to describe the
action of the Poisson brackets on these new classes of functions, that
are defined thanks to
formul{\ae}~\eqref{frm:esempio-g-in-PPset}--\eqref{frm:def-PPset}.
This is made as it follows.
\begin{lemma}
  \label{lem:classi-funzioni-x-tori-ellittici}
  Let us consider two generic functions $g\in\Pset_{\ell,sK}$ and
  $h\in\Pset_{m,rK}\,$, where $K$ is a fixed positive integer
  number. Then\footnote{The statement can be considered as valid also
    in the trivial cases with $\ell+m=0,1$, by enlarging the
    definition of the classes of functions so that
    $\Pset_{-2,sK}=\Pset_{-1,sK}=\big\{0\big\}\ \forall\>s\in\naturali$.},
  $$
  \big\{g,h\big\} = \Lie{h}\,g \in \Pset_{\ell+m-2,(r+s)K}
  \quad
  \ \forall\>\ell,\,m,\,r,\,s\in\naturali\ .
  $$
\end{lemma}
Also in this case the proof is omitted, because it can be obtained by
simply applying\footnote{Actually, it looks natural to be doubtful
  about the fact that Poisson brackets always preserve the
  restrictions on the Fourier harmonics that must be satisfied by
  variables of D'Alembert type. However, one can immediately realize
  that the only tricky case occurs when the Poisson brackets include
  also the following terms:
  $$
  \vcenter{\openup1\jot\halign{
      \hbox {\hfil $\displaystyle {#}$}
  &\hbox {$\displaystyle {#}$\hfil}\cr
  \frac{\partial \big(\sqrt{J_j})^{|\hat{\ell}_j|}\exp(\imunit \hat{\ell}_j\phi_j)}
  {\partial\phi_j}
  \frac{\partial \big(\sqrt{J_j})^{|\hat{m}_j|}\exp(\imunit \hat{m}_j\phi_j)}
       {\partial J_j}
  \cr\ -
  \frac{\partial \big(\sqrt{J_j})^{|\hat{\ell}_j|}\exp(\imunit \hat{\ell}_j\phi_j)}
  {\partial J_j}
  \frac{\partial \big(\sqrt{J_j})^{|\hat{m}_j|}\exp(\imunit \hat{m}_j\phi_j)}
       {\partial \phi_j}
  \cr
  \ \ =
  \frac{\imunit}{2}\big(\hat{\ell}_j|\hat{m}_j|-\hat{m}_j|\hat{\ell}_j|\big)
  \big(\sqrt{J_j})^{|\hat{\ell}_j|+|\hat{m}_j|-2}
  \exp\big(\imunit(\hat{\ell}_j+\hat{m}_j)\phi_j\big)\ .
  \cr
  }}
  $$
  However, if $\hat{\ell}_j$ and $\hat{m}_j$ have opposite signs
  then
  $\big|\hat{\ell}_j+\hat{m}_j\big|\le\big|\hat{\ell}_j\big|+\big|\hat{m}_j\big|-2$
  (let us remark that the term above vanishes if $\hat{\ell}_j=0$ or
  $\hat{m}_j=0$). In the remaining case (i.e.,
  $\hat{\ell}_j\neq 0$ and $\hat{m}_j\neq 0$ have the same sign), the
  coefficient $\hat{\ell}_j|\hat{m}_j|-\hat{m}_j|\hat{\ell}_j|$
  is always equal to zero.} the definition of the Poisson brackets.

As an environment where it is natural to properly define the algorithm
constructing the normal form for elliptic tori, let us start to
consider a Hamiltonian $\Hscr^{(0)}(\vet p, \vet q, \vet J, \vet\phi)$
that can be written in the following way:
\begin{equation}
\label{frm:espansione-H^(0)}
\vcenter{\openup1\jot 
\halign{
$\displaystyle\hfil#$&$\displaystyle{}#\hfil$&$\displaystyle#\hfil$\cr
  \Hscr^{(0)}(\vet p, \vet q, \vet J, \vet\phi) =
  & \Escr^{(0)}+\scalprod {\vet \omega^{(0)}}{\vet p }+
  \scalprod {\vet \Omega^{(0)}}{\vet J }+
  \sum_{s\ge 0}\sum_{\ell\ge 3} f_{\ell}^{(0,\,s)}(\vet p, \vet q, \vet J, \vet\phi)
  \cr
  & +
  \sum_{s\ge 1}\sum_{\ell =0}^2 f_{\ell}^{(0,\,s)}(\vet p, \vet q, \vet J, \vet\phi)
  \,,
  \cr
}}
\end{equation}
where $\Escr^{(0)}\in\Pset_{0,0}$ is a constant\footnote{$\Escr^{(0)}$
  denotes the energy level of the elliptic torus that is invariant in
  the approximation given by the angular average, i.e., when
  $f_{\ell}^{(0,\,s)}=0$ $\forall\ s>0$.} and $f_{\ell}^{(0,\,s)}\in
\Pset_{\ell,sK}$, being the first upper index related to the
normalization step. For instance, in~\cite{Car-Loc-2021} it is shown
how to bring an FPU chain of $N+1$ particles in the form above, by
following a procedure that is valid for a generic Hamiltonian problem
in the neighborhood of a stable equilibrium point. In other words, the
Hamiltonian describing that model can be expanded as $\Hscr^{(0)}$
in~\eqref{frm:espansione-H^(0)}, with $f_{\ell}^{(0,\,s)}=0$ when
$s\ge 3$ and $f_{\ell}^{(0,\,1)}\in\Pset_{\ell,K}\,$,
$f_{\ell}^{(0,\,2)}\in\Pset_{\ell,2K}$ $\forall\ \ell\ge 0$,
being\footnote{Setting $K=2$ is quite natural for Hamiltonian systems
  close to stable equilibria, see, e.g.,~\cite{Gio-Loc-San-2017}.}
$K=2$. This holds true, both for the so called $\alpha$--model and the
$\beta$ one. Let us also emphasize that the energy value
$\Escr^{(0)}$, the angular velocity vector
$\vet\Omega^{(0)}\in\reali^{n_2}$ and all the functions
$f_{\ell}^{(0,\,s)}$ depend on $\vet\omega^{(0)}\in\reali^{n_1}$ in a
parametric way. In order to keep the notation so that it does not get
too cumbersome, in the present subsection we do not include
$\vet\omega^{(0)}$ among the arguments of the terms appearing in the
expansions of the Hamiltonians. Moreover, for a generic problem in the
neighborhood of a stable equilibrium point one can also easily show
that $f_{\ell}^{(0,\,s)}=\Oscr(\epsilon^s)$, where $\epsilon$ is the
natural small parameter for this kind of models, because it denotes
the first approximation of the distance (expressed in terms of the
actions) between the wanted elliptic torus and the stable equilibrium
point.

In a strict analogy with what has been done to construct the
Kolmogorov normal form, here our main purpose is to eliminate from the
Hamiltonian all the terms having total degree less than three in the
square root of the actions; by referring to the paradigmatic form
described in~\eqref{frm:espansione-H^(0)}, the unwanted terms are
appearing in its last row. Actually, such a goal can be achieved by
performing an infinite sequence of canonical transformations, so as to
bring the Hamiltonian to the following final normal form:
\begin{equation}
\label{frm:espansione-H^(inf)}
\Hscr^{(\infty)}(\vet P, \vet Q, \vet\Xi, \vet\Theta) =
 \Escr^{(\infty)}+\scalprod {\vet \omega^{(\infty)}}{\vet p }+
 \scalprod {\vet \Omega^{(\infty)}}{\vet\Xi}
 +\sum_{s\ge 0}\sum_{\ell\ge 3}
 f_{\ell}^{(\infty,\,s)}(\vet P, \vet Q, \vet\Xi, \vet\Theta)\ ,
\end{equation}
with $f_{\ell}^{(\infty,\,s)}\in \Pset_{\ell,sK}$ and
$\Escr^{(\infty)}\in\Pset_{0,0}$. The motion law $( \vet P(t), \vet
Q(t), \vet\Xi(t), \vet\Theta(t)) = \big( \vet 0, \vet Q_0 + \vet
\omega^{(\infty)} t, \vet 0, \vet\Theta \big)$ is a solution of the
Hamilton equations related to the normal form $\Hscr^{(\infty)}$ and
is equivalent\footnote{We remark that $\dot{\vet P}=\big\{\vet
  P\,,\,\Hscr^{(\infty)}\big\}=\vet 0$ and
  $\dot{\vet\Xi}=\big\{\vet\Xi\,,\,\Hscr^{(\infty)}\vet\}=\vet 0$ when
  $\vet P=\vet 0$ and $\vet\Xi=\vet 0$. Because of the well known
  degeneracy of the change of coordinates $(\vet X,\vet
  Y)=\big(\sqrt{2\vet\Xi}\cos\vet\Theta,\sqrt{2\vet\Xi}\sin\vet\Theta\big)$,
  all the set $\big\{\big(\vet\Xi=\vet
  0,\vet\Theta\in\toro^{n_2}\big)\big\}$ correspond to a single point
  $\big\{(\vet X=\vet 0,\vet Y=\vet 0)\big\}$ of the reduced phase
  space that considers just the last $n_2$ degrees of freedom.  By the
  way, we emphasize that such a degeneracy is completely harmless in
  the framework we have adopted. In order to conclude the check of the
  solution of the Hamilton equations related to the normal form
  $\Hscr^{(\infty)}$ when $\vet P=\vet 0$ and $\vet\Xi=\vet 0$, it is
  enough to remark that $\dot{\vet Q}=\big\{\vet
  Q\,,\,\Hscr^{(\infty)}\big\}=\vet\omega^{(\infty)}$.}
to~\eqref{frm:soluzione-su-toro-ellittico}. Such a motion law is
generated by the initial condition $( \vet 0, \vet Q_0, \vet 0,
\vet\Theta)$, is quasi-periodic with an angular velocity vector equal
to $\vet\omega^{(\infty)}$ and the corresponding orbit lies on the
$n_1-$dimensional invariant torus $ \vet P= \vet 0$, $ \vet\Xi=\vet
0$. The energy level of such a manifold is $H^{(\infty)}(\vet 0, \vet
Q, \vet 0, \vet\Theta)=\Escr^{(\infty)}$. Moreover, it is elliptic in
the sense that the transverse dynamics in a neighborhood of the
invariant torus itself is given by oscillations whose corresponding
angular velocity vector is approaching $\vet{\Omega}^{(\infty)}$ in
the limit of $\|(\vet P,\vet\Xi)\|$ going to zero.

Also in the present case, that is concerning the elliptic tori, the
formal algorithm for the construction of the normal form is composed
by a sequence of canonical transformations, defined using the
formalism of Lie series.  We can summarize the $r$-th normalization
step, by giving the formula defining the canonical change of
coordinates that transforms the intermediate Hamiltonian
$\Hscr^{(r-1)}$ into the subsequent $\Hscr^{(r)}$. The expansion of the
former is of the following type:
\begin{equation}
\label{frm:espansione-H^(r-1)}
\vcenter{\openup1\jot 
\halign{
$\displaystyle\hfil#$&$\displaystyle{}#\hfil$&$\displaystyle#\hfil$\cr
\Hscr^{(r-1)}(\vet p, \vet q, \vet J, \vet\phi) &=
 &\Escr^{(r-1)}+\scalprod {\vet \omega^{(r-1)}}{\vet p }
 +\scalprod {\vet \Omega^{(r-1)}}{\vet J }+
 \sum_{s\ge 0}\sum_{\ell\ge 3} f_{\ell}^{(r-1,s)}(\vet p, \vet q, \vet J, \vet\phi)
 \cr
 & &+\sum_{s\ge r}\sum_{\ell =0}^2
 f_{\ell}^{(r-1,\,s)}(\vet p, \vet q, \vet J, \vet\phi)\ ,
 \cr
}}
\end{equation}
being $f_{\ell}^{(r-1,\,s)}\in \Pset_{\ell,sK}$ and
$\Escr^{(r-1)}\in\Pset_{0,0}\,$, i.e., it is a constant referring to
the level of the energy in the approximation that is valid up to terms
$\Oscr\big(\epsilon^r\big)$. Let us emphasize that the starting
Hamiltonian $\Hscr^{(0)}$ written in
equation~\eqref{frm:espansione-H^(0)} is exactly in the
form~\eqref{frm:espansione-H^(r-1)} with $r=1$. The conjugacy relation
which allows to write the Hamiltonian defined at the end of the $r$-th
normalization step as a function of the previous one is given by
\begin{equation}
  \label{frm:r-thnormstep}
  \Hscr^{(r)} = \Hscr^{(r-1)}\circ\exp\big(\Lie{\chi_0^{(r)}}\big)
  \circ \exp\big(\Lie{\chi_1^{(r)}}\big)\circ
  \exp\big(\Lie{\chi_2^{(r)}}\big) \circ \Dgot^{(r)}\ ,
\end{equation}
where the Lie series\footnote{Because of the so called ``exchange
  theorem'' (see~\cite{Grobner-60}), the new Hamiltonian $H^{(r)}$ is
  obtained from the old one, by applying the Lie series to $H^{(r-1)}$
  in reverse order with respect to what is written
  in~\eqref{frm:r-thnormstep}. This is consistent with the order of
  the discussion in the following subsections: the first stage of the
  $r$-th normalization step deals with the canonical transformation
  generated by $\chi_0^{(r)}$, the second one with $\chi_1^{(r)}$ and
  the last one with both $\chi_2^{(r)}$ and $\Dgot^{(r)}$.}
operator $\exp\big(\Lie{\chi_j^{(r)}}\big) \,\cdot$ removes the
Hamiltonian terms with total degree in the square root of the actions
equal to $j$ and with trigonometric degree in the angles $\vet q$ up
to $rK$.  Moreover, by a linear canonical transformation
$\Dgot^{(r)}$, the terms that are quadratic in $\sqrt{\vet J}$ and do
not depend on both the actions $\vet p$ and the angles $\vet q$ are
brought to a diagonal form. At the end of this $r$-th normalization
step, the ineliminable terms that are independent on the angles $\vet
q$ and linear either in $\vet p$ or in $\vet J$ are added to the
normal form part. This requires to update the angular velocities from
$\big( \vet \omega^{(r-1)}, \vet \Omega^{(r-1)}\big)$ to $\big( \vet
\omega^{(r)}, \vet \Omega^{(r)}\big)$, that is why
in~\eqref{frm:espansione-H^(inf)} the Hamiltonian in Kolmogorov normal
form has new frequency vectors $\vet \omega^{(\infty)}$ and
$\vet\Omega^{(\infty)}$.

All the details that properly define how the algorithm actually works
are exhaustively described in the following.

\subsubsection{First stage of the $r$-th normalization step.\\}\label{sss:first-stage}
In the context of the $r$-th normalization step, the first stage aims
to remove the terms depending just on the angles $\vet q$ up to the
trigonometrical degree $rK$, i.e. the terms collected in
$f_{0}^{(r-1,r)}=\Oscr(\epsilon^r)$.  We determine the generating
function $\chi^{(r)}_{0}$ by solving the homological equation
\begin{equation}
\poisson{\scalprod{ \vet \omega^{(r-1)}}{ \vet p}}{\chi^{(r)}_{0}}
+ f_{0}^{(r-1,\,r)}(\vet q) = \langle f_{0}^{(r-1,r)}(\vet q) \rangle_{\vet q}\ .
\label{frm:chi0r}
\end{equation}
Let us remark that the equation above is perfectly equivalent to that
in formula~\eqref{frm:chi1rKAM}, because
$f_{0}^{(r-1,\,r)}\in\Pset_{0,rK}$ depends on $\vet q$ only and,
therefore, $f_{0}^{(r-1,\,r)}\in\Pscr_{0,rK}=\Pset_{0,rK}\,$.  Thus,
we can write the solution of this new (first) homological
equation~\eqref{frm:chi0r} exactly in the same way as we have done for
what concerns~\eqref{frm:espansione-chi1rKAM}, i.e., we put $\langle
f_{0}^{(r-1,r)}(\vet q) \rangle_{\vet q}=c_{\vet 0}$ and
\begin{equation}
\chi^{(r)}_{0}( \vet q)=\sum_{{\scriptstyle{0<| \vet k|\le rK}}}
  \frac{c_{\vet k} \exp\big(\imunit\scalprod{\vet k}{ \vet q}\big)}
  {\imunit\scalprod{\vet k}{\vet\omega^{(r-1)}}}\ ,
\label{frm:espansione-chi0r}
\end{equation}
being $f_{0}^{(r-1,\,r)}( \vet q)=\sum_{{\scriptstyle{| \vet k|\le rK}}}
c_{\vet k} \exp\big(\imunit\scalprod{\vet k}{ \vet q}\big)$.
Of course, such a solution is certainly valid provided the
non-resonance
condition~\eqref{frm:nonres-cond-passor-KAM-primomezzopasso} is
satisfied.

Now, we apply the canonical transformation $\exp\lie{\chi_0^{(r)}}$ to
the Hamiltonian which is defined at the end of the $r-1$-th
normalization step. By the usual abuse of notation, we choose to
rename the new variables as the old ones. This allows to write the
transformed Hamiltonian $H^{({\rm I}; \,
  r)}=\exp\big(\Lie{\chi^{(r)}_{0}}\big)H^{(r-1)}$ as follows:
\begin{equation}
\vcenter{\openup1\jot 
\halign{
$\displaystyle\hfil#$&$\displaystyle{}#\hfil$&$\displaystyle#\hfil$\cr
\Hscr^{({\rm I}; \, r)}(\vet p, \vet q, \vet J, \vet\phi) &=
 &\Escr^{(r)}+\scalprod {\vet \omega^{(r-1)}}{\vet p }
 +\scalprod {\vet \Omega^{(r-1)}}{\vet J }+
 \sum_{s\ge 0}\sum_{\ell\ge 3} f_{\ell}^{({\rm I};\,r,\,s)}
 \cr
 & &+\sum_{s\ge r}\sum_{\ell =0}^2
 f_{\ell}^{({\rm I}; \,r,\,s)}\ ,
 \cr
}}
\label{frm:H(I;r)-espansione}
\end{equation}
where for the sake of brevity we have omitted to list the arguments of
the functions $f_{\ell}^{({\rm I}; \,r,\,s)}$.  Let us introduce them
in the same unconventional way we have adopted in
Subsection~\ref{sbs:Kolmog-original} to describe the algorithm
constructing the Kolmogorov normal form.  First, we define\footnote{We
  remark that the terms $f_{\ell}^{(r-1,\,s)}$ do not enter in the
  expansion~\eqref{frm:espansione-H^(r-1)} when $\ell=0,\,1,\,2$ and
  $s<r$. However, the recursive definitions described in the present
  subsection are such that all those functions are equal to
  zero. Keeping in mind this fact allows to write in a rather compact
  way both formula~\eqref{frm:ridefinizioni-fI} and the analogous ones
  in the following.} $f_{\ell}^{({\rm I};
  \,r,\,s)}=f_{\ell}^{(r-1,\,s)}$ $\forall\ \ell\ge 0,\,s\ge 0$.  By
further abuses of notation, we update many times the definition of the
terms appearing in the expansion of the new Hamiltonian according to
the following rule:
\begin{equation}
\label{frm:ridefinizioni-fI}
  f_{\ell -2i}^{({\rm I}; \,r,\,s+jr)} \hookleftarrow
  \frac{1}{j!} \lie {\chi_0^{(r)}}^j f_{\ell}^{(r-1,\,s)}
  \quad \forall \  \ell \ge 0, \ 1\le j \le \lfloor \ell/2 \rfloor,
  \ s \ge 0\ .
\end{equation}
By applying repeatedly
Lemma~\ref{lem:classi-funzioni-x-tori-ellittici} and a trivial
induction argument to the formula above, one can easily prove that
$f_{\ell}^{({\rm I}; \,r,\,s)}\in \Pset_{\ell,sK}$ $\forall\ \ell\ge
0,\,s\ge 0$. In order to end the description of the first stage of the
$r$-th normalization step, we have to take into account also the
effects induced by the homological equation~\eqref{frm:chi0r}. For
such a purpose, we finally set $f_{0}^{({\rm I}; \,r,\,r)}=0$ and we
update the approximated value referring to the energy of the wanted
elliptic torus exactly in the same way we have done to write
formula~\eqref{frm:ridefinizione-utile-x-energia}, i.e., we put
$\Escr^{(r)}=\Escr^{(r-1)}+\langle f_{0}^{(r-1,\,r)}\rangle_{\vet
  q}\,$.

\subsubsection{Second stage of the $r$-th normalization step.\\}
\label{sss:second-stage}
The second stage of the $r$-th normalization step acts on the
Hamiltonian that is initially expanded as
in~\eqref{frm:H(I;r)-espansione}, with the goal to remove the
perturbing term which is linear in $\sqrt{\vet J}$ and independent of
$\vet p$, i.e., $f_1^{({\rm I}; \,r,\,r)}$. Thus, we have to solve the
following homological equation:
\begin{equation}
\label{frm:chi1r}
\poisson{\scalprod{\vet \omega^{(r-1)}}{\vet p}
          + \scalprod{\vet \Omega^{(r-1)}}{\vet J}}{\chi_1^{(r)}}
          + f_{1}^{({\rm I}; \,r,\,r)}(\vet q, \vet J, \vet\phi)=0\ .
\end{equation}
Let us write the expansion of $f_{1}^{({\rm I}; \,r,\,r)}(\vet q, \vet
J, \vet \phi)$ as follows:
\begin{equation}
\label{frm:espansione-f_1^(I,r,r)}
f_{1}^{({\rm I}; \,r,\,r)} (\vet q, \vet J, \vet \phi) =
\sum_{0\le \vet k\le rK} \sum_{j=1}^{n_2} \sqrt{J_j}
\left[ c_{\vet k, \,j}^{\scriptscriptstyle(+)}\,e^{\imunit(\scalprod{\vet k}{\vet q}+\phi_j)}
  + c_{\vet k, \,j}^{\scriptscriptstyle(-)}\,e^{\imunit(\scalprod{\vet k}{\vet q}-\phi_j)}\right]\ ,
\end{equation}
where every coefficients $c_{\vet k, \,j}^{\scriptscriptstyle(+)}\in\complessi$ is equal
to the complex conjugate of $c_{-\vet k, \,j}^{\scriptscriptstyle(-)}$ $\forall\ 0\le
\vet k\le rK$, $1\le j\le n_2\,$. Therefore, the generating function
$\chi_1^{(r)}$ solving equation~\eqref{frm:chi1r} is determined in
such a way that
\begin{equation}
\label{frm:espansione-chi1r}
\chi_1^{(r)}(\vet q, \vet J, \vet \phi) =
\sum_{0\le \vet k\le rK} \sum_{j=1}^{n_2} \frac{\sqrt{J_j}}{\imunit}
\left[
  \frac{c_{\vet k, \,j}^{\scriptscriptstyle(+)}\,e^{\imunit(\scalprod{\vet k}{\vet q}+\phi_j)}}
       {\scalprod{\vet k}{\vet \omega^{(r-1)}}+\Omega_j^{(r-1)}}
  + \frac{c_{\vet k, \,j}^{\scriptscriptstyle(-)}\,e^{\imunit(\scalprod{\vet k}{\vet q}-\phi_j)}}
         {\scalprod{\vet k}{\vet \omega^{(r-1)}}-\Omega_j^{(r-1)}}
\right]\ .
\end{equation}
This expression is well-defined, provided that the frequency vector
$\vet \omega^{(r-1)}$ satisfies the so-called first Melnikov
non-resonance condition up to order $rK$ (see~\cite{Melnikov-1965}),
i.e.,
\begin{equation}
\label{frm:non-resIIstep}
\min_{{0 <|\vet k|\le rK,}\atop {|\vet \ell |= 1}}
\left|\scalprod{\vet k}{\vet \omega^{(r-1)}}
+ \scalprod{\vet \ell} {\vet \Omega^{(r-1)}}
\right|\ge \frac{\gamma}{(rK)^\tau}
\quad{\rm and}\quad
\min_{|\vet \ell| =1} \left| \scalprod{\vet \ell}{\vet \Omega^{(r-1)}}\right|
\ge \gamma\,,
\end{equation}
for some fixed values of both $\gamma>0$ and $\tau>n_1-1$.  By
applying the Lie series $\exp\big(\lie{\chi_1^{(r)}}\big)$ to the old
Hamiltonian $H^{({\rm I};\,r)}$, we have a new one, which we denote as
$H^{({\rm II};\,r)}=\exp\big(\Lie{\chi^{(r)}_{1}}\big)H^{({\rm
    I};\,r)}$ and have the same structure as that described
in~\eqref{frm:H(I;r)-espansione}, i.e.,
\begin{equation}
\vcenter{\openup1\jot 
\halign{
$\displaystyle\hfil#$&$\displaystyle{}#\hfil$&$\displaystyle#\hfil$\cr
\Hscr^{({\rm II}; \, r)}(\vet p, \vet q, \vet J, \vet\phi) &=
 &\Escr^{(r)}+\scalprod {\vet \omega^{(r-1)}}{\vet p }
 +\scalprod {\vet \Omega^{(r-1)}}{\vet J }+
 \sum_{s\ge 0}\sum_{\ell\ge 3} f_{\ell}^{({\rm II};\,r,\,s)}
 \cr
 & &+\sum_{s\ge r}\sum_{\ell = 0}^2
 f_{\ell}^{({\rm II}; \,r,\,s)}\ ,
 \cr
}}
\label{frm:H(II;r)-espansione}
\end{equation}
The functions $f_{\ell}^{({\rm II}; \,r,\,s)}$ that compose the new
Hamiltonian can be determined with calculations similar to those
listed during the description of the first stage of normalization.
This means that we initially define $f_{\ell}^{({\rm II};
  \,r,\,s)}=f_{\ell}^{({\rm I}; \,r,\,s)}$ $\forall\ \ell\ge 0,\,s\ge
0$. Then, (by abuse of notation) we redefine them many times according
to the following rules:
\begin{equation}
  \label{frm:ridefinizioni-fII}
  \vcenter{\openup1\jot 
    \halign{
      $\displaystyle\hfil#$&$\displaystyle{}#\hfil$&$\displaystyle#\hfil$\cr
  f_{\ell-j}^{({\rm II}; \,r,\,s+jr)} &\hookleftarrow
  &\frac{1}{j!} \lie {\chi_1^{(r)}} ^j  f_{\ell}^{({\rm I}; \,r,\,s)} 
  \quad \forall \ \ell \ge 0\,, \ 1 \le j \le \ell\,, \ s \ge 0 \ ,
  \cr
  f_{0}^{(r, \,2r)} &\hookleftarrow
  &\frac{1}{2}\lie{\chi_1^{(r)}}^2 \big(\scalprod {\vet \omega^{(r)}}{\vet p }
  +\scalprod {\vet \Omega^{(r)}}{\vet J }\big)\ .
  \cr
  }}
\end{equation}
Because of the homological equation~\eqref{frm:chi1r}, we add also a
further redefinition so that $f_{1}^{({\rm II}; \,r,\,r)}=0$.  By
applying Lemma~\ref{lem:classi-funzioni-x-tori-ellittici} to
formula~\eqref{frm:ridefinizioni-fII}, it is easy to check that
$f_{\ell}^{({\rm II}; \,r,\,s)}\in \Pset_{\ell,sK}$ $\forall\ \ell\ge
0,\,s\ge 0$.

\subsubsection{Third stage of the $r$-th normalization step.\\}
\label{sss:third-stage}
The third and last stage of normalization is more elaborated. It aims
to remove terms belonging to two different classes: first, those
linear in $\vet p$ and independent of $(\vet J,\vet\phi)$, moreover,
other terms that are quadratic in $\sqrt{\vet J}$ and independent of
$\vet p$.  Such a part of the perturbation is removed by the
composition of two canonical transformations expressed by Lie series,
being the corresponding generating functions $X_2^{(r)}(\vet p, \vet
q)\in\PPset_{1,0,rK}$ and $Y_2^{(r)}(\vet q, \vet
J,\vet\phi)\in\PPset_{0,2,rK}$, respectively. Moreover, the third
stage is ended by a linear canonical transformation $\Dgot^{(r)}$
that leaves the pair $(\vet p,\vet q)$ unchanged and it aims to
diagonalize the terms that are quadratic in $\sqrt{\vet J}$ and
independent of the angles $\vet q$. Let us detail all these changes of
coordinates, so that the algorithm will be unambiguously defined at
the end of our discussion.

The generating functions $X_2^{(r)}$ is in charge to remove terms that
are linear in $\vet p$ and do depend on the angles $\vet q$ up to
the trigonometric degree $rK$. Therefore, it is a solution of the
following homological equation:
\begin{equation}
\label{frm:X2r}
\poisson{\scalprod{\vet \omega^{(r-1)}}{\vet p}}{X_2^{(r)}}
+ f_{2}^{({\rm II};\,r,\,r)}(\vet p, \vet q)
- \langle f_{2}^{({\rm II}; \,r,\,r)}(\vet p, \vet q) \rangle_{\vet q} = 0 \ .
\end{equation}
Let us recall that $f_{2}^{({\rm II}; \,r,\,r)}\in\Pset_{2,rK}=
\PPset_{1,0,rK}\,\cup\,\PPset_{0,2,rK}\,$; indeed, such a function
does depend on all the canonical variables, i.e., $f_{2}^{({\rm II};
  \,r,\,r)}= f_{2}^{({\rm II}; \,r,\,r)}(\vet p,\vet q,\vet
J,\vet\phi)$.  Therefore, we denote with $f_{2}^{({\rm II};
  \,r,\,r)}(\vet p,\vet q)$ the subpart of $f_{2}^{({\rm II};
  \,r,\,r)}$ that is depending just on $(\vet p,\vet q)$. Analogously,
in the following $f_{2}^{({\rm II}; \,r,\,r)}(\vet q,\vet J,\vet\phi)$
will denote the subpart of $f_{2}^{({\rm II}; \,r,\,r)}$ that does
depend on all the canonical variables but the actions $\vet p$ and so
on also for what concerns $f_{2}^{({\rm II};\,r,\,r)}(\vet
J,\vet\phi)$. For the sake of clarity, this highly non-standard
notation will be maintained up to the end of the present subsection.
Let us here emphasize that the term $\langle f_{2}^{({\rm II};
  \,r,\,r)}(\vet p, \vet q) \rangle_{\vet q}$ will be added to the
part in normal form, by updating the angular velocity vector
$\vet\omega$, in agreement with what has been done in the context of
the construction of the Kolmogorov normal form.  We can deal with the
homological equation~\eqref{frm:X2r} in the same way as
for~\eqref{frm:chi2rKAM}. Indeed, the solution writes as
\begin{equation}
\label{frm:espansione-X2r}
  X_2^{(r)}(\vet p, \vet q) = 
  \sum_{{\scriptstyle{|\vet j|=1}}}\,
  \sum_{{\scriptstyle{0<| \vet k|\le rK}}}
  \frac{c_{\vet j,\vet k}\, {\vet p}^{\vet j}
    \exp\big(\imunit\scalprod{\vet k}{ \vet q}\big)}
  {\imunit\scalprod{\vet k}{\vet\omega^{(r)}}}\ ,
\end{equation}
where the expansion of the perturbing term $f_{2}^{({\rm II};
  \,r,\,r)}(\vet p,\vet q)\in\PPset_{1,0,rK}$ is such that $f_{2}^{({\rm
    II}; \,r,\,r)}(\vet p,\vet q)= \sum_{{\scriptstyle{|\vet j|=1}}}\,
\sum_{{\scriptstyle{0<| \vet k|\le rK}}} c_{\vet j,\vet k}\, {\vet
  p}^{\vet j} \exp\big(\imunit\scalprod{\vet k}{ \vet q}\big)$.  Once
again, the solution written in~\eqref{frm:espansione-X2r} is
valid provided that the non-resonance
condition~\eqref{frm:nonres-cond-passor-KAM-primomezzopasso} is
satisfied.

The generating function $Y_2^{(r)}$ aims to remove the part of the
term of $f_{2}^{({\rm II}; \,r,\,r)}$ that is quadratic in
$\sqrt{\vet J}$ and does depend on the angles $\vet q$. Therefore,
$Y_2^{(r)}$ has to solve the following homological equation:
\begin{equation}
\label{frm:Y2}
\poisson{\scalprod{\vet \omega^{(r-1)}}{\vet p}
  + \scalprod{\vet \Omega^{(r-1)}}{\vet J}}{Y_2^{(r)}}
+ f_{2}^{({\rm II}; \,r,\,r)}(\vet q,\vet J, \vet\phi)
- \langle f_{2}^{({\rm II}; \,r,\,r)} (\vet q,\vet J, \vet\phi)\rangle_{\vet q}
= 0 \ .
\end{equation}
In order to describe the solution of such an equation, it is
convenient to write the explicit expansion of the perturbing term
$f_{2}^{({\rm II}; \,r,\,r)} (\vet q,\vet J,\vet\phi)$. For instance,
this can be done in the following way:
\begin{equation}
\label{frm:espansione-f_2^(I,r,r)}
  f_{2}^{({\rm II};\, r,\,r)} (\vet q, \vet J, \vet \phi) =
  \sum_{0\le \vet k\le rK} \sum_{i,\,j=1}^{n_2} c_{\vet k, \,i,\,j}^{\scriptscriptstyle(\pm, \pm)}
    \,\sqrt{J_i J_j}
    \exp\big[\imunit(\scalprod{\vet k}{\vet q} \pm \phi_i \pm \phi_j)\big]\ ,
\end{equation}
where $c_{\vet k, \,i,\,j}^{\scriptscriptstyle(+,+)}$ and $c_{\vet k, \,i,\,j}^{\scriptscriptstyle(+,-)}$
are the coefficients referring to the Fourier harmonics
$\scalprod{\vet k}{\vet q}+\phi_i+\phi_j$ and $\scalprod{\vet k}{\vet
  q}+\phi_i-\phi_j\,$, respectively, and so on.  Thus, the generating
function $Y_2^{(r)}$ is determined by equation~\eqref{frm:Y2} in such
a way that
\begin{equation}
  \label{frm:espansione-Y2r}
  Y_2^{(r)}(\vet q, \vet J, \vet \phi) =
  \sum_{0< \vet k\le rK} \sum_{i,\,j=1}^{n_2}
  \frac{ c_{\vet k, \,i,\,j}^{\scriptscriptstyle(\pm, \pm)}\,\sqrt{J_i J_j}
    \exp\big[\imunit(\scalprod{\vet k}{\vet q} \pm \phi_i \pm \phi_j)\big]}
       {\imunit\big(\scalprod{\vet k}{\vet \omega^{(r-1)}}\pm
         \Omega_i^{(r-1)}\pm\Omega_j^{(r-1)}\big)}
    \ ,
\end{equation}
which is well defined provided that the angular velocity vector $\vet
\omega^{(r-1)}$ satisfies both the already mentioned Diophantine
inequality~\eqref{frm:nonres-cond-passor-KAM-primomezzopasso} and the
so-called second Melnikov non-resonance condition up to order $rK$
(see~\cite{Melnikov-1965}), i.e.,
\begin{equation}
\label{frm:non-resIIIstep}
\min_{{0 <|\vet k|\le rK,}\atop{  |\vet \ell| = 2 }}
\left|\scalprod{\vet k}{\vet \omega^{(r-1)}}+
\scalprod{\vet \ell}{\vet \Omega^{(r-1)}} \right|\ge \frac{\gamma}{(rK)^\tau}
\end{equation}
with fixed values of both parameters $\gamma>0$ and $\tau>n_1-1$.

After having performed these two changes of coordinates, we still may
have terms that do not depend on $\vet q$ and are either linear in
$\vet p$ or quadratic in $\sqrt{\vet J}$. The former ones can be
directly added to the part in normal form, whereas the latter have to
be preliminarily put in diagonal form. This can be done by means of a
canonical transformation $\Dgot^{(r)}$ such that
\begin{equation}
\label{frm:def-diagonalizza}
\left(\scalprod{\vet\Omega^{(r-1)}}{\vet J}
+f_{2}^{({\rm II};\,r,\,r)}(\vet J,\vet\phi)\right)
\Bigg|_{(\vet J,\vet\phi)=\Dgot^{(r)}(\bar{\vet J},\bar{\vet\phi})}
=
\scalprod{\vet\Omega^{(r)}}{\bar{\vet J}}\ .
\end{equation}
Such an equation in the unknown transformation $\Dgot^{(r)}$
can be solved provided that
\begin{equation}
\label{frm:non-resIII2}
\min_{|\vet \ell| = 2 }
\left|\scalprod{\vet \ell}{\vet \Omega^{(r-1)}} \right|
\ge \gamma
\end{equation}
and $f_{2}^{({\rm II};\,r,\,r)}$ is small enough, as it is explained,
e.g., in section~7 of~\cite{Gio-et-al-89} (where this problem is
considered in the equivalent case dealing with polynomial canonical
coordinates). In practical implementations, such a change of
coordinates $\Dgot^{(r)}$ can be conveniently defined by composing a
subsequence of Lie series, each of them being related to a quadratic
generating function $\Dscr_2^{(r;\,m)}(\vet
J,\vet\phi)\in\PPset_{0,2,0}$ with $m\in\naturali\setminus\{0\}$.  All
these new generating functions can be determined by adopting the
following computational (sub)procedure of iterative type.  First, we
introduce the new angular velocity vector $\vet\Omega^{(r;\,0)}$ so
that
\begin{equation}
  \label{frm:def-Omega-r;0}
  \scalprod{\vet\Omega^{(r;\,0)}}{\vet J} =
  \scalprod{\vet\Omega^{(r-1)}}{\vet J}
  +\langle f_{2}^{({\rm II};\,r,\,r)}(\vet J,\vet\phi)\rangle_{\vet\phi}
\end{equation}
and the new function
\begin{equation}
  \label{frm:def-g2-r;0}
  \ggot_{2}^{(r;\,0)}(\vet J,\vet\phi) =
  f_{2}^{({\rm II};\,r,\,r)}(\vet J,\vet\phi) -
  \langle f_{2}^{({\rm II};\,r,\,r)}(\vet J,\vet\phi)\rangle_{\vet\phi}\ .
\end{equation}
The general $m$-th step of this iterative (sub)procedure starts by
solving the following homological equation:
\begin{equation}
\label{frm:D2-r;m}
\poisson{\scalprod{\vet\Omega^{(r;\,m-1)}}{\vet J}}
        {\Dscr_2^{(r;\,m)}(\vet J,\vet\phi)}
+ \ggot_{2}^{(r;\,m-1)}(\vet J,\vet\phi) = 0 \ ,
\end{equation}
where $\ggot_{2}^{(r;\,m-1)}\in\PPset_{0,2,0}$ is such that
$\langle\ggot_{2}^{(r;\,m-1)}\rangle_{\vet\phi}=0$ (and, therefore,
also the new generating function $\Dscr_2^{(r;\,m)}$ is sharing these
same properties with $\ggot_{2}^{(r;\,m-1)}$). Let us now initially
introduce $\ggot_{2}^{(r;\,m)}=0$ and (by the usual
abuse of notation) we redefine it many times according to the
following rule:
\begin{equation}
  \label{frm:ridefinizioni-g2}
  \ggot_{2}^{(r;\,m)} \hookleftarrow
  \frac{j}{(j+1)!}\lie{\Dscr_2^{(r;\,m)}}^j \ggot_{2}^{(r;\,m-1)}
  \quad \forall \ j \ge 1\ .
\end{equation}
Actually, at this point one can easily check that
$$
\exp\big(\lie{\Dscr_2^{(r;\,m)}}\big)\,
\big(\scalprod{\vet\Omega^{(r;\,m-1)}}{\vet J}+
\ggot_{2}^{(r;\,m-1)}\big)=
\scalprod{\vet\Omega^{(r;\,m-1)}}{\vet J}+\ggot_{2}^{(r;\,m)}\ ,
$$
by using homological equation~\eqref{frm:D2-r;m}. Furthermore,
we set
\begin{equation}
  \label{frm:def-Omega-r;m}
  \scalprod{\vet\Omega^{(r;\,m)}}{\vet J} =
  \scalprod{\vet\Omega^{(r;\,m-1)}}{\vet J}
  +\langle \ggot_{2}^{(r;\,m)}(\vet J,\vet\phi)\rangle_{\vet\phi}
\end{equation}
and we redefine one last time $\ggot_{2}^{(r;\,m)}$ so that
\begin{equation}
  \label{frm:ultima-ridef-g2-r;m}
  \ggot_{2}^{(r;\,m)}(\vet J,\vet\phi) =
  \ggot_{2}^{(r;\,m)}(\vet J,\vet\phi) -
  \langle \ggot_{2}^{(r;\,m)}(\vet J,\vet\phi)\rangle_{\vet\phi}\ .
\end{equation}
By applying repeatedly
Lemma~\ref{lem:classi-funzioni-x-tori-ellittici} to
formul{\ae}~\eqref{frm:def-Omega-r;0}--\eqref{frm:ultima-ridef-g2-r;m},
it is easy to check that both functions $\Dscr_2^{(r;\,m)}$ and
$\ggot_{2}^{(r;\,m)}$ belong to the class $\PPset_{0,2,0}$ (also
because they depend on neither $\vet p$ nor $\vet q$) and their
angular average is equal to zero. In principle, these remarks would
allow to iterate infinitely many times this computational
(sub)procedure, that we are using to solve
equation~\eqref{frm:def-diagonalizza}.  However, in practical
implementations, we have to set a criterion to stop the iterations so
to ensure that the algorithm can be worked out in a finite number of
operations. This can be done, for instance, in such a way to end the
computations when the angular velocity vector does not modify
anymore. This means that the final value $\bar m$ of the normalization
step for this iterative (sub)procedure is such that the equation
$\vet\Omega^{(r;\,{\bar m})}=\vet\Omega^{(r;\,{\bar m}-1)}$ holds true
\emph{in the framework of the numbers that are representable on a
  computer}\footnote{A similar criterion is adopted to determine a
  maximum value of the index $j$ at which the
  redefinitions~\eqref{frm:ridefinizioni-g2} must be stopped.} (for
instance, the {\tt double precision} type). By setting
$\vet\Omega^{(r)}=\vet\Omega^{(r;\,{\bar m})}$ and the canonical
transformation $\Dgot^{(r)}$ equal to composition of all the Lie
series generated by the \emph{finite} sequence of functions
$\big\{\Dscr_2^{(r;\,m)}\big\}_{m=1}^{\bar m}$, we determine a
solution\footnote{As an alternative computational method, when one is
  dealing with the estimates needed to prove the convergence of the
  algorithm, in~\cite{Gio-Loc-San-2014} the use of the Lie transforms
  (that are equivalent to the composition of {\it infinite} sequences
  of Lie series) has been found to be very suitable.}
of~\eqref{frm:def-diagonalizza} that is valid up to the numerical
round-off errors.

Finally, we need to understand how all these generating functions
(that have been defined during the third stage of the $r$-th
normalization step) give their contributions to the Hamiltonian terms
appearing in the following expansion:
\begin{equation}
\label{frm:espansione-H^(r)}
\vcenter{\openup1\jot 
\halign{
$\displaystyle\hfil#$&$\displaystyle{}#\hfil$&$\displaystyle#\hfil$\cr
\Hscr^{(r)}(\vet p, \vet q, \vet J, \vet\phi) &=
 &\Escr^{(r)}+\scalprod {\vet \omega^{(r)}}{\vet p }
 +\scalprod {\vet \Omega^{(r)}}{\vet J }+
 \sum_{s\ge 0}\sum_{\ell\ge 3} f_{\ell}^{(r,s)}(\vet p, \vet q, \vet J, \vet\phi)
 \cr
 & &+\sum_{s\ge r+1}\sum_{\ell =0}^2
 f_{\ell}^{(r,\,s)}(\vet p, \vet q, \vet J, \vet\phi)\ ,
 \cr
}}
\end{equation}
where $\Hscr^{(r)}$ is defined in~\eqref{frm:r-thnormstep}.  In order
to describe the definitions of those new summands, it is convenient to
introduce the intermediate functions $g_{\ell}^{(r, \,s)}$,
${g'}_{\ell}^{(r, \,s)}$ in the following way. First, we define
$g_{\ell}^{(r, \,s)} = f_{\ell}^{({\rm II};\, r, \,s)}$ for all
non-negative values of the indexes $\ell$ and $s$; then, we consider
the effects induced by the application of the Lie series with
generating function $X_2^{(r)}$ to the Hamiltonian. In order to do
that, (by abuse of notation) we redefine many times the new
intermediate functions $g_{\ell}^{(r, \,s)}$ according to the
following rules:
\begin{equation}
  \label{frm:ridefinizioni-g}
  \vcenter{\openup1\jot 
    \halign{
      $\displaystyle\hfil#$&$\displaystyle{}#\hfil$&$\displaystyle#\hfil$\cr
  g_{\ell}^{(r, \,s+jr)} &\hookleftarrow
  &\frac{1}{j!}\lie{X_2^{(r)}}^j f_{\ell}^{({\rm II};\, r, \,s)}
  \quad \forall \ j \ge 1,\, \ell \ge 0, \, s \ge 0\ ,
  \cr
  g_{2}^{(r, \,jr)} &\hookleftarrow
  &\frac{1}{j!}\lie{X_2^{(r)}}^j \big(\scalprod {\vet \omega^{(r)}}{\vet p }
  +\scalprod {\vet \Omega^{(r)}}{\vet J }\big)
  \quad \forall \ j \ge 1\ .
  \cr
  }}
\end{equation}
As usual, the prescriptions above have been set so to gather the new
terms generated by the Lie series $\exp\big(\lie{X_2^{(r)}}\big)$
according to both their total degree in the square root of the actions
and the trigonometric degree in the angles. In analogous way, we first
introduce ${g'}_{\ell}^{(r, \,s)}={g'}_{\ell}^{(r, \,s)}$
$\forall\ \ell\ge 0,\> s\ge 0$; then we apply many times the following
redefinitions:
\begin{equation}
  \label{frm:ridefinizioni-g'}
  \vcenter{\openup1\jot 
    \halign{
      $\displaystyle\hfil#$&$\displaystyle{}#\hfil$&$\displaystyle#\hfil$\cr
  {g'}_{\ell}^{(r, \,s+jr)} &\hookleftarrow
  &\frac{1}{j!}\lie{Y_2^{(r)}}^j g_{\ell}^{(r, \,s)}
  \quad \forall \ j \ge 1,\, \ell \ge 0, \, s \ge 0\ ,
  \cr
  {g'}_{2}^{(r, \,jr)} &\hookleftarrow
  &\frac{1}{j!}\lie{Y_2^{(r)}}^j \big(\scalprod {\vet \omega^{(r)}}{\vet p }
  +\scalprod {\vet \Omega^{(r)}}{\vet J }\big)
  \quad \forall \ j \ge 1\ .
  \cr
  }}  
\end{equation}
By applying Lemma~\ref{lem:classi-funzioni-x-tori-ellittici} to
formul{\ae}~\eqref{frm:ridefinizioni-g}--\eqref{frm:ridefinizioni-g'},
it is easy to check that ${g'}_{\ell}^{(r,\,s)}\in \Pset_{\ell,sK}$
$\forall\ \ell\ge 0,\,s\ge 0$.  Let us now remark that each class of
type $\Pset_{\ell,sK}$ is preserved\footnote{This statement can be
  justified, by referring also to the definition of the canonical
  transformation $\Dgot^{(r)}$ as composition of all the Lie series
  generated by the set of functions
  $\big\{\Dscr_2^{(r;\,m)}\big\}_{m=1}^{\bar m}$. In fact, it can be
  easily done by applying
  Lemma~\ref{lem:classi-funzioni-x-tori-ellittici} to all the
  contributions due to the repeated application of the Lie derivative
  with generating functions $\Dscr_2^{(r;\,m)}\in\Pset_{2,0}$.} by the
diagonalization transformation $\Dgot^{(r)}$, for all non-negative
values of the indexes $\ell$ and $s$. Therefore, it is natural to put
\begin{equation}
  f_{\ell}^{(r,s)} = {g'}_{\ell}^{(r, \,s)}\circ\Dgot^{(r)}\ .
\end{equation}
for all indexes $\ell\ge 0$ and $s\ge 0$.

At the end of the $r$-th normalization step, it is convenient that
the terms linearly depending just on $\vet p$ or $\vet J$ are included
in the main part of the Hamiltonian, because all of them belong to the
same class of functions, i.e. $\Pset_{2,0}$. For this purpose, we
introduce the new angular velocity vector $\vet \omega^{(r)}$, in such
a way that
\begin{equation}
\label{frm:nuove-freq}
\scalprod{\vet \omega^{(r)}}{\vet p} =
\scalprod{\vet \omega^{(r-1)}}{\vet p} + f_{2}^{({\rm II};\, r,\, 0)}(\vet p)\ ,
\end{equation}
while the new values of the components of $\vet\Omega^{(r)}$ are
defined by equation~\eqref{frm:def-diagonalizza}, that also allows us
to put $f_{2}^{(r,r)}=0$. This ends the justification of the fact that
the Hamiltonian $\Hscr^{(r)}$ can be written as in
formula~\eqref{frm:espansione-H^(r)} with new terms such that
$f_{\ell}^{(r,\,s)}\in \Pset_{\ell,sK}$ and
$\Escr^{(r)}\in\Pset_{0,0}\,$. Therefore, $\Hscr^{(r)}$ has the same
structure of $\Hscr^{(r-1)}$ in~\eqref{frm:espansione-H^(r-1)}; this
also mean that the normalization algorithm can be iterated to the next
($r+1$-th) step. As a final comment ending the present subsection, let
us also remark that the new perturbative terms $f_{\ell}^{(r,s)}$ with
$\ell = 0, \,1,\,2$ are expected to be smaller with respect to the
previous ones; this is because of the Fourier decay of the
coefficients jointly with the fact that we removed the part of
perturbation up to the trigonometric degree $rK$.

\subsection{On the convergence of the algorithm constructing the normal form for elliptic tori}
\label{subsec:theorem}

As we have discussed since the introduction, in the present work we
make the choice of adopting the same approach to construct two
different normal forms, that are related to KAM invariant manifolds
and elliptic tori, respectively. For what concerns the analysis of the
convergence, such a choice now allows us to use arguments that are very
similar to those described in the previous
Section~\ref{sec:KAMbasics}. In particular, also for what concerns the
motion on elliptic tori, we emphasize that it can be approximated
within a precision up to a fixed order of magnitude by using our
procedure that is \emph{explicitly computable}, because the total
amount of operations that are defined also by this normalization
algorithm is \emph{finite}.

The non-resonance conditions we have assumed
in~\eqref{frm:nonres-cond-passor-KAM-primomezzopasso},
\eqref{frm:non-resIIstep}, \eqref{frm:non-resIIIstep}
and~\eqref{frm:non-resIII2} can be summarized in the following way:
\begin{equation}
\label{frm:non-res-tot}
\min_{{0 <|\vet k|\le rK,}\atop {0\le |\vet \ell |\le2 }}
\left|\scalprod{\vet k}{\vet \omega^{(r-1)}}+
\scalprod{\vet \ell}{\vet\Omega^{(r-1)}}\right|\ge \frac{\gamma}{(rK)^\tau}
\quad{\rm and}\quad
\min_{0< |\vet \ell| \le2}
\left| \scalprod{\vet \ell}{\vet \Omega^{(r-1)}}\right|\ge \gamma\ ,
\end{equation}
with $\gamma>0$ and $\tau>n_1-1$.  Let us here resume the parametric
dependence of all the Hamiltonian terms on the initial value of the
angular velocity vector $\vet \omega^{(0)}$, as it has been introduced
at the beginning of the previous
Subsection~\ref{sbs:Kolmog-4-ell-tori} (see the discussion following
the statement of Lemma~\ref{lem:classi-funzioni-x-tori-ellittici}). In
particular, in the Diophantine inequalities reported
in~\eqref{frm:non-res-tot} the angular velocity vectors at the $r$-th
normalization step are functions of $\vet \omega^{(0)}$, i.e.,
$\vet\omega^{(r-1)}=\vet\omega^{(r-1)}(\vet\omega^{(0)})$ and
$\vet\Omega^{(r-1)}=\vet\Omega^{(r-1)}(\vet\omega^{(0)})$.  Let us
recall that we do not try to keep a full control on the way for what
concerns the angular velocity vectors that are modified passing from
the $r-1$-th normalization step to the next one. Therefore, let us
recall also here that such an approach is in contrast with the
original proof scheme that was designed to construct the Kolmogorov
normal form for {\it maximal} invariant tori, where the angular
velocities are kept fixed (see~\cite{Kolmogorov-1954} or,
e.g.,~\cite{Gio-Loc-1997}), but it is somehow unavoidable because of
the occurrence of the transversal angular velocities
$\vet\Omega^{(r-1)}(\vet\omega^{(0)})$ that in general cannot remain
constant along the normalization procedure. This seems to prevent the
complete construction of the normal form and so also for what concerns
the proof of the existence of an elliptic torus. Nevertheless,
following the approach designed by P{\"o}schel in~\cite{Poschel-1989},
it can be proved that the Lebesgue measure of the resonant regions
where the Melnikov conditions are not satisfied shrinks to zero with
the size of the perturbation. Therefore, the chances of success in
constructing the normal form for elliptic tori are described by the
following statement.

\begin{theorem}
  \label{unico-teorema}
Consider the family of real Hamiltonians $\Hscr^{(0)}(\vet p, \vet
q,\vet J, \vet\phi ;\vet\omega^{(0)})$ of the type described
in~\eqref{frm:espansione-H^(0)}. Those functions are defined so that
$\Hscr^{(0)}:\,\Oscr_1\times\toro^{n_1}\times\Oscr_2\times\toro^{n_2}\times\Uscr\mapsto\reali$,
with $\Oscr_1$ and $\Oscr_2$ open neighborhoods of the origin in
$\reali^{n_1}$ and $\reali_+^{n_2}\cup\{\vet 0\}$, respectively, while
$\vet \omega^{(0)}\in \Uscr$, being $\Uscr$ an open subset of
$\reali^{n_1}$.  Moreover, let a special class of functions include
each of the terms that are of type $f_{\ell}^{(0,\,s)}$ and appear in
the expansion~\eqref{frm:espansione-H^(0)}, in such a way that
$f_{\ell}^{(0,\,s)}\in\Pset_{\ell,sK}$ for a fixed positive integer
$K$. We also assume that

\item{(a)} all the functions $\Escr^{(0)}:\Uscr\mapsto\reali$,
  $\vet\Omega^{(0)}:\Uscr\mapsto\reali^{n_1}$ and
  $f_{\ell}^{(0,\,s)}:\Oscr_1\times\toro^{n_1}\times\Oscr_2\times\toro^{n_2}\times\Uscr\mapsto\reali$,
  appearing in~\eqref{frm:espansione-H^(0)}, are analytic functions with
  respect to $\vet\omega^{(0)}\in\Uscr$;

\item{(b)} $\Omega_i^{(0)}(\vet \omega^{(0)})\neq\Omega_j^{(0)}(\vet\omega^{(0)})$
  and $\Omega_{i_2}^{(0)}(\vet \omega^{(0)})\neq 0$ for
  $\vet \omega^{(0)}\in\Uscr$ and $1\le i<j\le n_2$, $1\le i_2\le n_2\,$;

\item{(c)} for some fixed and positive values of $\epsilon$ and $E$, one has
\begin{equation}
  \sup_{(\vet p,\vet q,\vet J,\vet\phi;\vet\omega^{(0)})\in
    \Oscr_1\times\toro^{n_1}\times\Oscr_2\times\toro^{n_2}\times\Uscr}
  \left|f_\ell^{(0,s)}(\vet p,\vet q,\vet J,\vet\phi;\vet\omega^{(0)})\right|
  \le \epsilon^s E
 \end{equation}
$\forall\ s\ge 1,\ \ell\ge 0$ and $\forall\ \ell\ge 3$ when $s=0$.

\noindent
Then, there is a positive $\epsilon^{\star}$ such that for $0\le
\epsilon<\epsilon^{\star}$ the following statement holds true: there
exists a non-resonant set $\,\Uscr^{(\infty)}\subset\Uscr$ of positive
Lebesgue measure and with the measure of
$\,\Uscr\setminus\Uscr^{(\infty)}$ tending to zero for $\epsilon\to 0$
for bounded $\Uscr$, such that for each $\vet
\omega^{(0)}\in\Uscr^{(\infty)}$ there exists an analytic canonical
transformation $(\vet p,\vet q,\vet J,\vet\phi)= \psi_{\epsilon;\vet
  \omega^{(0)}}^{(\infty)}(\vet P,\vet Q,\vet\Xi,\vet\Theta)$ leading
the Hamiltonian to the normal form written
in~\eqref{frm:espansione-H^(inf)}, where
$\Escr^{(\infty)}(\vet\omega^{(0)})$ is a finite real value fixing the
constant energy level that corresponds to the invariant elliptic torus $\big\{(
\vet P = \vet 0,\, \vet Q\in\toro^{n_1}, \vet\Xi=\vet 0,
\vet\Theta=\vet 0)\big\}\,$.  Moreover, the canonical change of
coordinates is close to the identity in the sense that
$\big\|\psi_{\epsilon;\vet\omega^{(0)}}^{(\infty)}(\vet P, \vet
Q,\vet\Xi,\vet\Theta)-(\vet P, \vet
Q,\vet\Xi,\vet\Theta)\big\|=\Oscr(\epsilon)$ and the same applies also
to both the energy level and the detunings of the angular velocity
vectors (that are
$\big|\Escr^{(\infty)}(\vet\omega^{(0)})-\Escr^{(0)}(\vet\omega^{(0)})\big|=\Oscr(\epsilon)$,
$\big\|\vet\omega^{(\infty)}(\vet\omega^{(0)})-\vet\omega^{(0)}\big\|=\Oscr(\epsilon)$
and
$\big\|\vet\Omega^{(\infty)}(\vet\omega^{(0)})-\vet\Omega^{(0)}(\vet\omega^{(0)})\big\|=\Oscr(\epsilon)$,
respectively).
\end{theorem}
The complete proof of theorem above is reported
in~\cite{Caracciolo-2021}, where it is ensured the convergence of a
normalization algorithm that is substantially the same with respect to
the one described in the previous
Subsection~\ref{sbs:Kolmog-4-ell-tori} apart some very minor
modifications\footnote{For instance, in order to describe the
  transverse dynamics with respect to the elliptic tori, the complex
  canonical coordinates $(\vet z,\imunit\bar{\vet z})$ instead
  of the action-angle ones are used, where $z_j=J_j
  e^{\imunit\phi_j}$ $\forall\ j=1,\,\ldots\,,\,n_2\,$.}.  Therefore,
the approach of that paper is based on a convergence scheme of linear
type. Nevertheless, the more geometrical part of that work (which
deals with the estimates of the volume covered by the resonant
region) is borrowed from~\cite{Poschel-1989}, where a statement nearly
equivalent to Theorem~\ref{unico-teorema} is proved by adopting a fast
convergence scheme of quadratic type.

In the present case studying the elliptic tori, the choice to let the
angular velocity vectors change at every normalization step is somehow
more natural with respect to the original proof scheme designed by
Kolmogorov. This is due to the fact that here the procedure allowing
to keep fixed the angular velocities is not complete, because it
involves less free parameters than the number of degrees of
freedom. This is a major difference with respect to the algorithm
constructing the normal form for KAM tori, where those two integer
numbers are equal. For what concerns the case of the elliptic tori
too, some work\footnote{Danesi, V., Locatelli, U.: work in progress
  (2021).} is in progress in order to revisit the problem of the
convergence of this type of normalization algorithms so as to provide
a statement where the final result is not expressed in a probabilistic
sense (i.e., by referring to the Lebesgue measure). This can be done
by fixing since the beginning the final value of the angular velocity
vectors $\big(\vet\omega^{(\infty)},\vet\Omega^{(\infty)}\big)$ and
their non-resonance properties; we emphasize that this allow to
explicitly solve all the homological equations that are introduced at
every step of the algorithm.  Also here, the total detunings
$\vet\omega^{(\infty)}-\vet\omega^{(0)}$ and
$\vet\Omega^{(\infty)}-\vet\Omega^{(0)}$ are given in terms of series
whose coefficients are defined in a recursive way.  Such an approach
is also inspired by the need to revisit what was successfully done in
order to show the existence of elliptic tori in PDEs problems
(see~\cite{Ber-Bia-2011}).

\section{Construction of invariant KAM tori in exoplanetary systems with rather eccentric orbits}\label{sec:exotori}

In order to properly introduce a Cauchy problem { which includes
  the ordinary differential equations (ODE) for a planetary system},
the initial conditions at a given time are needed and so also for the
positions and the velocities in an astrocentric frame. It is well
known that they can be replaced by the orbital elements
$$
\Big\{ \big(a_j\,,\,e_j\,,\,\iota_j\,,\,M_j\,,\,\omega_j\,,\,\Omega_j\big)\,:
\ \forall\ j=1,\,\ldots\,,N\Big\}\ ,
$$ being $N$ the number of the planets that are considered in the
system.  Orbital elements refer to the so called osculating Keplerian
ellipse, which describes a fictitious motion having the same
instantaneous values of both position and velocity with respect to the
planet.  For what concerns the Keplerian ellipse of the $j$-th planet,
the symbols
$a_j\,,\,e_j\,,\,\iota_j\,,\,M_j\,,\,\omega_j\,,\,\Omega_j$ denote the
semi-major axis, the eccentricity, the inclination\footnote{
  $\iota_j$ is the inclination of the Keplerian ellipse with respect
  to the plane orthogonal to the line of sight (i.e., the direction
  pointing to the object one is observing), that is usually said to be
  ``tangent to the celestial sphere''.}, the mean anomaly, the
argument of the pericenter\footnote{Unfortunately, the same symbol
  (namely, $\omega$) is used to denote both the angular velocity in
  KAM theory and the pericenter argument in astronomy. Hereafter, when
  the symbol $\omega$ appears \emph{without} superscripts, it will
  refer just to the latter quantity.} and the longitude of the
ascending node, respectively. Of course, also the values of the masses
$m_j$ $\forall\ j=0,\,1,\,\ldots\,,\,N$ (being $m_0$ the stellar mass)
are needed in order to properly introduce the Cauchy problem for a
planetary system, because they enter in the definitions of the
momenta, the kinetic energy and the potential one. Unfortunately, none
of the detection methods that are nowadays available to discover
extrasolar planets is able to measure all the orbital elements and the
masses that completely define the ODE problem (see,
e.g.,~\cite{Beau-FerM-Mich-2012}). For the sake of simplicity, instead
of considering a generic planetary problem with $N+1$ bodies, let us
focus on a specific case, i.e., the extrasolar system hosting two
planets orbiting around the star named HD$\,$4732\footnote{ Since
  the detection of a fainter stellar companion in 2019 (see
  \cite{Mug-2019}) HD$\,$4732 has been renamed as HD$\,$4732A. For
  brevity, in the present paper we refer to such a star with the old
  name.}  (the value of its mass is reported in the caption of the
following table). The values of the known orbital elements of those
exoplanets as they are given by the radial velocity detection method
are reported in Table~\ref{tab:RV-measures-orb-el}. Let us recall that
such a detection technique is unable to provide a complete information
about the mass of every $j$-th planet; instead, it gives its minimum
value $m_j\sin(\iota_j)$.

\begin{table}
  \caption{Known orbital elements and minimal masses of the detected
    exoplanets orbiting around the HD$\,$4732 star, whose mass
    is~$1.74$ times bigger than the solar one. The following data are
    taken from the central values of the ranges given in Table~5
    of~\cite{Sato-et-al-2013}. The corresponding units of measure are
    reported in every column between pairs of square brackets; in
    particular, we recall that the eccentricity of an ellipse is a
    pure number ranging in $(0,1)$ and ${\rm M}_{\rm Jup}$ means
    ``Jupiter mass''. Since the initial time is irrelevant for an
    autonomous system, we have set it equal to zero in the parentheses
    following the orbital elements.}
\begin{center}
\begin{tabular}{c c c c c c}
  \hline
  Planet &$\quad$Planet$\quad$ &$\quad a_j(0)\quad$ & $\quad e_j(0)\quad$ & $\quad\omega_j(0)\quad$ &$\quad m_j\sin\big(\iota_j(0)\big)\quad$\\
  name &index $j$ &[AU]  &       & $[{{\circ}\atop\phantom{1}}]$ &$[{\rm M}_{\rm Jup}]$\\
  \hline
  HD$\,$4732b &1 & 1.19 & 0.13 & 85 & 2.37 \\
  HD$\,$4732c &2 & 4.60 & 0.23 & 118 & 2.37 \\
  \hline
\end{tabular}
\end{center}
  \label{tab:RV-measures-orb-el}
\end{table}

Let us now explain how we have decided to complete the initial
conditions, by also giving the motivations of our choice. Since we are
interested in studying the planetary dynamics of the HD$\,$4732 system
in the framework of a secular model, we expect that its dependence on
the initial values of the mean anomalies is weak. {We emphasize
  that such an assumption does not hold true in general (see,
  e.g.,~\cite{Lib-San-2013}), but it is rather natural in the case of
  the HD$\,$4732 planetary system because the revolution periods are
  far from mean-motion resonances and they are }much shorter with
respect to those corresponding to the remaining angles that appear in
the orbital elements list. Therefore, we simply set\footnote{{
    Since the times of passage at the pericenter are given by the
    radial velocity detection methods and they are different, we
    stress that our choice of defining the initial values of the mean
    anomalies so that $M_1(0)=M_2(0)=0^{\circ}$ is not coherent with
    the observations about the two planets orbiting around
    HD$\,$4732. However, we consider that this small inconsistency of
    our settings should be harmless, just because of the expectation
    that its secular dynamics should be very weakly affected by the
    initial values of the mean anomalies.}}
\begin{equation}
  \label{def:M-iniz}
  M_1(0)=M_2(0)=0^{\circ}\ .
\end{equation}
For what concerns the extrasolar system HD$\,$4732, we plan to start a
study of the dependence of its orbital dynamics on the mutual
inclination $i_{\rm mut}\,$. The present section deals with the
beginning of such a research project, that will be extended in a
forthcoming work. For this purpose, it is convenient to consider
orbital planes initially located in such a way they are symmetric with
respect to the line of sight that is also orthogonal to their
intersection.  As an example of this particular configuration, we can
consider the case with $\iota_1(0)=89^{\circ}$,
$\iota_2(0)=91^{\circ}$ and
\begin{equation}
  \label{def:Omega-iniz}
  \Omega_1(0)=\Omega_2(0)=0^{\circ}\ .
\end{equation}
In view of the general relation
$$
\cos i_{\rm mut} = \cos \iota_1 \cos \iota_2 +
\sin \iota_1 \sin \iota_2 \cos(\Omega_1-\Omega_2)\ ,
$$
we readily obtain that $i_{\rm mut}=2^\circ$. More in general,
we introduce the following set of initial conditions 
\begin{equation}
  \label{def:set-cond-iniz}
  \vcenter{\openup1\jot 
    \halign{
      $\displaystyle\hfil#$&$\displaystyle{}#\hfil$&$\displaystyle#\hfil$\cr
      \Iscr_{i_{\rm mut}(0)} &=
      \Big\{ \big( &a_1(0)\,,a_2(0)\,,\,e_1(0)\,,\,e_2(0)\,,
      \cr
      & &\iota_1(0)=90^{\circ}-\frac{i_{\rm mut}(0)}{2}\,,\,
      \iota_2(0)=90^{\circ}+\frac{i_{\rm mut}(0)}{2}\,,\,
      \cr
      & & M_1(0)\,,\,M_2(0)\,,\omega_1(0)\,,\,\omega_2(0)\,,
      \Omega_1(0)\,,\Omega_2(0)\big)\Big\}\ ,
      \cr
  }}
\end{equation}
where the inclinations are parameterized with respect to $i_{\rm
  mut}(0)$, while the values of all the remaining orbital elements are
defined according to Table~\ref{tab:RV-measures-orb-el}, jointly with
formul{\ae}~\eqref{def:M-iniz} and~\eqref{def:Omega-iniz}.  Of course,
the values of the planetary masses $m_1$ and $m_2$ can be recovered
multiplying the minimal masses (that appear in the last column of
Table~\ref{tab:RV-measures-orb-el}) by the increasing factor $1{\big
  /}\sin\big(\iota_j(0)\big)$. This remark helps us to understand that
all the parameters and the initial contidions have been properly
defined and they can eventually depend just on the value of $i_{\rm
  mut}(0)$. This way to parameterize the model has been introduced to
better understand the properties of our (new) algorithm constructing
invariant tori as a function of the mutual inclinations.  A previous
approach to the same problem was described in~\cite{Vol-Loc-San-2018}
and it was shown to be successful just for systems with rather small
eccentricities of the exoplanets, being their initial values less than
$0.1\,$. This is not the case of the exoplanets in the system
HD$\,$4732, because both their initial values of the eccentricities
(reported in Table~\ref{tab:RV-measures-orb-el}) are larger than
$0.1\,$. We emphasize that this choice has been made with the purpose
to show that our following new formulation of the constructing
algorithms applies to a more extended range of models with respect to
the previous approach.

Let us also recall that, in a three-body planetary problem, the
longitudes of the nodes are always opposite, if they are measured with
respect to the so called Laplace plane, that is invariant because it
is orthogonal to the total angular momentum, by definition {(see,
  e.g., section~6.2 of~\cite{Laskar-1989})}. Moreover, the Hamiltonian
does not depend on the sum of $\Omega_1+\Omega_2\,$, because of the
invariance with respect to the rotations.  In the following
subsection, we will explain why it is preferable to consider
expansions of the Hamiltonian in a frame where the Laplace plane is
the horizontal one. In Celestial Mechanics the word ``inclination''
often refers to the angle (say, $\igot_j\in[0^{\circ},180^{\circ}]$)
between the angular momentum of the $j$-th planet and the total
one. With this notation, the following relation holds true: $i_{\rm
  mut} = \igot_1+\igot_2\,$.

\subsection{Secular model at order two in the masses}
\label{sec:secular-model}
In the present subsection, we are going to introduce a model
describing the secular dynamics of a planetary system, in a way that
provides results more reliable with respect to a simple average over
the revolution angles (see, e.g.,~\cite{San-Lib-2019}). We emphasize
that we derive the secular model at order two in the masses, by
applying an approach inspired to the construction of the Kolmogorov
normal form. This is a major difference with respect to other approaches
providing the same level of accuracy for a secular model (see,
e.g.,~\cite{Laskar-1988} and references therein). Here, in order to
introduce our secular model, we will adopt the approach described
in~\cite{Vol-Loc-San-2018}, that is summarized as follows.

A three-body Hamiltonian problem has nine degrees of freedom, but
three of them can be easily separated so as to describe the uniform
motion of the center of mass in an inertial frame. The untrivial part
of the dynamics is represented in astrocentric canonical coordinates
and its degrees of freedom can be further reduced by two using the
conservation of the total angular momentum $\vet C$. As it is shown in
section~6 of~\cite{Laskar-1989}, this allows us to write the
Hamiltonian in Poincar\'e canonical variables, that are
\begin{equation}
  \label{frm:poin-var}
  \vcenter{\openup1\jot 
    \halign{
      $\displaystyle\hfil#$&$\displaystyle{}#\hfil$&$\displaystyle#\hfil$
      &$\displaystyle\hfil#$&$\displaystyle{}#\hfil$&$\displaystyle#\hfil$\cr
      \Lambda_j &= &\frac{m_0 m_j}{m_0 + m_j} \sqrt{G(m_0+m_j)a_j}\ ,\quad
      &\xi_j & = &\sqrt{2 \Lambda_j}
      \sqrt{1-\sqrt{1-e_j^2}}\cos{(\omega_j)}\ ,
      \quad
      \cr
      \lambda_j &= & M_j + \omega_j\ ,
      &\eta_j &= &-\sqrt{2 \Lambda_j} \sqrt{1-\sqrt{1-e_j^2}}\sin{(\omega_j)}
      \ .
      \cr
  }}
\end{equation}
The reduction of the total angular momentum makes implicit the
dependence on the inclinations $\igot_j$ and on the longitudes of the
nodes $\Omega_j\,$. In the Laplace reference frame the mutual
inclination is the sum of the two inclinations and so is given by a
rather simple relation involving the Poincar\'e variables, i.e.,
\begin{equation}
  i_{\rm mut} =
  \igot_1 +\igot_2 =
  \arccos\left(\frac{C^2-\Lambda_1^2(1-e_1^2)-\Lambda_2^2(1-e_2^2) }
              {2\Lambda_1 \Lambda_2 \sqrt{1-e_1^2}\sqrt{1-e_2^2}}\right)\ ,
\end{equation}
being $C = \sum_{k=1}^2 \Lambda_k \sqrt{1-e_k^2}\cos \igot_k\,$, that
is the (constant) module of the total angular momentum.  Moreover, we
introduce a translation $L_j = \Lambda_j - \Lambda_j^*$, where
$\Lambda_j^*$ is defined in order to obtain that in the Keplerian
approximation of the motion the values of the semi-major axes are in
agreement with the observations. Indeed, the expansions of a
Hamiltonian representing a planetary model are usually made around the
average values of the semi-major axes or their initial values. For the
sake of simplicity, we will adopt this latter option. Such expansions
are actually made with respect to these Poincar\'e
variables\footnote{The computation of the coefficients appearing in
  the expansion~\eqref{frm:3BP} is not straightforward. For a detailed
  discussion of the method we have used for doing such a calculation
  we refer to~\cite{Laskar-1989}.} and the parameter $D_2$, that
measures the difference between the total angular momentum of the
system and the one of a similar system with circular and coplanar
orbits; i.e., it is defined as $D_2=\big[(\Lambda_1^*+\Lambda_2^*)^2
  -C^2\big]\big/(\Lambda_1^* \Lambda_2^*)$; therefore, it is of the
same order as $e_1^2+\igot_1^2+e_2^2+\igot_2^2\,$.  Thus, we can write
the Hamiltonian of the three--body problem as
\begin{equation}
 \label{frm:3BP}
  H_{\rm 3BP} = \sum_{j_1=1}^{\infty}h^{({\rm Kep})}_{j_1,0}(\vet{L}) +
  \mu\sum_{s=0}^{\infty}\sum_{j_1=0}^{\infty}\sum_{j_2=0}^{\infty}\,D_2^s\,
  h^{(\mathcal{P})}_{s;j_1,j_2}(\vet{L},\boldsymbol{\lambda},\boldsymbol{\xi},\boldsymbol{\eta})
\end{equation}
where $\mu = \max\{m_1/m_0,m_2/m_0\}$. Moreover,
\begin{itemize}
\item $ \Kscr(\vet L)= \sum_{j_1=1}^{\infty}h^{({\rm Kep})}_{j_1,0}(\vet{L})$
  is the Keplerian part and $h^{({\rm Kep})}_{j_1,0}$ is a
  homogeneous polynomial of degree $j_1$ in $\vet{L}$; in
  particular, $h^{({\rm Kep})}_{1,0} = \vet{n}^*\cdot\vet{L}$, where the
  components of the angular velocity vector $\vet{n}^*$ are defined by
  the third Kepler law;
\item $h^{(\Pscr)}_{s;j_1,j_2}$ is a homogeneous polynomial of degree
  $j_1$ in $\vet{L}$, degree $j_2$ in
  $(\boldsymbol{\xi},\boldsymbol{\eta})$ and with coefficients that
  are trigonometric polynomials in $\vet\lambda$ and are related to
  the term $D_2^s\,$.
\end{itemize}
Clearly, in the applications we deal with finite expansions; the
truncation parameters will be discussed in the following.

The expression of the Hamiltonian of the three-body problem
in~\eqref{frm:3BP} highlights the distinction between the so called
{\it fast variables} $(\vet L, \vet \lambda)$ and the {\it secular
  variables} $(\vet \xi, \vet \eta)$. Indeed, if we consider the
corresponding Hamilton equations, we have that $\dot {\vet \lambda} =
\Oscr(1)$. This means that the motion of the planet along the orbit,
that is in first approximation a Keplerian ellipse, has a different
timescale with respect to the secular variables, whose variation is
due to the interaction between the planets and, therefore, is of
$\Oscr(\mu)$. Since we are interested in the study of the long-time
stability of the system, a common procedure consists on considering
just the evolution of the secular variables, by averaging the
Hamiltonian with respect to the fast angles $ \vet \lambda$. With a
simple average of $H_{\rm 3BP}$ we would obtain a secular
approximation with terms of order $\mu$, namely at order~$1$ in the
masses. Here, we consider terms up to order $2$ in the masses,
averaging with a close to the identity canonical change of coordinates
inspired by the algorithm for the construction of the Kolmogorov
normal form. Indeed, we focus on the torus corresponding to
$\boldsymbol{L}=0$.  The first transformation of coordinates that we
define aims at removing the perturbative terms that depend on the
angles $\vet \lambda$ but do not depend on the actions $\vet L$, being
$\dot{L_j}=\partial H/\partial\lambda_j$ for $j=1,2\,$.  This is done
by using the term linear in the actions, i.e., $\vet n^* \cdot \vet
L$, to define a generating function $\chi^{(\Oscr 2)}_1(\vet \lambda)$
as the solution of the following homological equation:
\begin{equation}
  \label{eq:homeq}
  \left\{\chi^{(\Oscr 2)}_1,\ \vet n^* \cdot \vet L\right\} +
  \mu \sum_{{s=0\,,\>j_2=0}\atop{2s+j_2\le
      N_S}}\left\lceil D_2^s\, h_{s;0,j_2}^{(\Pscr)}\right\rceil_%
      {\boldsymbol{\lambda}:K_F}= \mu
      \sum_{{s=0\,,\>j_2=0}\atop{2s+j_2\le N_S}}D_2^s\, \Big\langle
      h_{s;0,j_2}^{(\Pscr)}\Big\rangle_{\boldsymbol{\lambda}}\ ,
\end{equation}
being $\langle \cdot \rangle_{\vet \lambda}$ the average with respect
to the angles $\vet \lambda$, while with the notation $\lceil \cdot
\rceil_{K_F}$ we mean that the expansions are truncated at the
trigonometrical degree $K_F$ in the angles $\vet \lambda$.  Let us add
a few comments about the truncations parameters $K_F$ and $N_S$. The
value of $K_F$ is defined so as to take into account the main
mean-motion quasi-resonances of the system considered.  For example,
if the system is close to the resonance $k_1^*:k_2^*$, then $K_F$ is
defined as $K_F \ge |k_1^*|+|k_2^*|$. In the same spirit, the value
$N_S$ of the truncation of the expansions in eccentricity and
inclination is set in order to consider the quasi-resonance. Let us
assume that the quasi-resonant angular terms are of type
$(k_1^*\lambda_1 - k_2^* \lambda_2)$, then in principle it would be
convenient to consider expansions up to an order in eccentricity and
inclination such that $N_S\ge 2(|k_1^*|-|k_2^*|)$, because of the
D'Alembert rules (see~\cite{Laskar-1989}). Therefore, in the specific
case of the extrasolar system HD$\,$4732, it is rather natural to set
$K_F = 9$, because the periods of the two planets are about
$0.986\,$yr and $7.48$\,yr, respectively. However, since the ratio of
the angular velocities $n_1^*/n_2^*$ is not so close to the resonance
$7:1$ or to $8:1$ and the terms of high degree in eccentricities are
not so relevant, we have found convenient to limit our expansions to
$N_S=8\,$, in order to reduce the computational cost of the whole
procedure.

Now we have to apply the transformation of coordinates defined by the
application of the Lie series operator $\exp(\Lie{\chi_1^{(\Oscr
    2)}})\ \cdot = \sum_{j=0}^\infty (1/j!)\Lie{\chi_1^{(\Oscr
    2)}}^j\,\cdot$ to the Hamiltonian. Recalling that in our secular
model we will not consider terms depending on $\vet L$ or of order
greater than $\mu^2$, the only terms we need to compute are included
in the following expansion:
\begin{equation}
  \label{eq:htilde}
  \vcenter{\openup1\jot 
    \halign{
      $\displaystyle\hfil#$&$\displaystyle#\hfil$\cr
    &\widetilde{\kern-2pt H} = H_{\rm 3BP} +
    \frac{1}{2}\left\{\chi^{(\Oscr 2)}_1,
    \ \Lscr_{\chi^{(\Oscr  2)}_1}h^{({\rm Kep})}_{2,0}\right\}_{\vet L, \vet \lambda}
    \cr
    &\>+\mu \, \sum_{{s\ge 0\,,\>j_2\ge 0}\atop{2s+j_2\le N_S}} D_2^s
    \left\{\chi^{(\Oscr 2)}_1, {h}_{s;1,j_2}^{(\Pscr)}\right\}_{\vet L, \vet \lambda}
    + \frac{\mu}{2}\,\sum_{{s\ge 0\,,\>j_2\ge 0}\atop{2s+j_2\le N_S}} D_2^s
    \left\{\chi^{(\Oscr 2)}_1,{h}_{s;0,j_2}^{(\Pscr)}\right\}_{\vet \xi,\vet \eta}\ ,
    \cr
  }}
\end{equation}
where $\fastpoisson{\cdot}{\cdot}$ and $\secpoisson{\cdot}{\cdot}$ are
the terms of the Poisson bracket involving only the derivatives with
respect to the pairs of conjugate variables $(\vet L,\vet \lambda)$
and $(\vet \xi,\vet \eta)$, respectively. Then, according
to~\cite{Loc-Gio-2000}, we have that
\begin{equation*}
  \langle H^{(\Oscr 2)} \rangle_{\vet{\lambda}} \Big|_{\vet{L}= \vet{0}}
  = \langle\, \widetilde{\kern-2pt H}\, \rangle_{\vet{\lambda}}
  \Big|_{\vet{L}= \vet{0}} + {\cal O} (\mu^3)\,,
\end{equation*}
being $H^{(\Oscr 2)} = \exp(\Lie{\chi_1^{(\Oscr2)}})H_{\rm 3BP}\,$.
Let us remark that for the definition of this model it is not
necessary to compute the effects induced by the second generating
function $\chi_2^{(\Oscr_2)}(\vet L,\vet \lambda)$ for removing terms
linear in $\vet L$, because the additional terms due to the
application of such a Lie series operator are neglected in the secular
approximation.

We can finally introduce our secular model up to order $2$ in the
masses by setting
\begin{equation}
  \label{eq:hamsec}
  H^{({\rm sec})}(D_2,\boldsymbol{\xi},\boldsymbol{\eta}) =
  \left\lceil\,
  \langle\,\widetilde{\kern-2pt H}\,\rangle_{\vet{\lambda}}
  \Big|_{\vet{L}= \vet{0}}\,\right\rceil_{N_S}\, ,
\end{equation}
i.e., we take the averaged expansion (over the fast angles
$\vet{\lambda}$) of the part of $\,\widetilde{\kern-2pt H}$ that is
both independent from the actions $\vet L$ and truncated up to a total
order of magnitude $N_S$ in eccentricity and inclination. Since $D_2$
is $\Oscr\big(e_1^2+\igot_1^2+e_2^2+\igot_2^2\big)$, this means that
we keep the Hamiltonian terms $h^{(\Pscr)}_{s;0,j_2}$ with $2s+j_2\le
N_S$. From now on, the parameter $D_2$ is replaced by its explicit
value that is calculated as a function of the initial conditions;
thus, we can write the Hamiltonian as follows:
\begin{equation}
\label{frm:hsec}
H^{({\rm sec})}(\vet \xi,\vet \eta) =
\sum_{s=1}^{N_S/2} h^{({\rm sec})}_{2s}(\vet \xi,\vet \eta)\ ,
\end{equation}
where $h_{2s}$ is an homogeneous polynomial of degree $2s$. This means
that the expansion contains just terms of even degree, as a further
consequence of the well known D'Alembert rules. To fix the ideas, in
the case of the extrasolar system HD$\,$4732 let us emphasize that our
secular model at order two in the masses is defined by a Hamiltonian
$H^{({\rm sec})}$ that is a simple (even) polynomial of maximal
degree~8 in the four canonical variables $(\vet \xi,\vet \eta)$.

We have explicitly performed all the computations of Poisson brackets
(required by Lie series formalism to express canonical
transformations) and all the expansions described in the present
subsection and in in the next one, by using {\it X$\rho$\'o$\nu
  o\varsigma$}. It is a software package especially designed for doing
computer algebra manipulations into the framework of Hamiltonian
perturbation theory (see~\cite{Gio-San-Chronos-2012} for an
introduction to its main concepts).

\subsection{Semi-analytic computations of invariant tori}
\label{sec:normal-form-4-tori}

In the framework of Hamiltonian theory for dynamical systems, often
intuition can be fruitfully helped by numerical investigations.  In
particular, in the case of the extrasolar system HD$\,$4732, they
allow to easily motivate the new approach that is based on normal
forms and we are going to describe. In the present section, we will
discuss some results provided by direct numerical integrations of the
secular model $H^{({\rm sec})}$ that is defined in~\eqref{frm:hsec};
all of them have been produced by simply applying the RK4 method.

\begin{figure}
\centering
  \includegraphics[width=6.0cm]{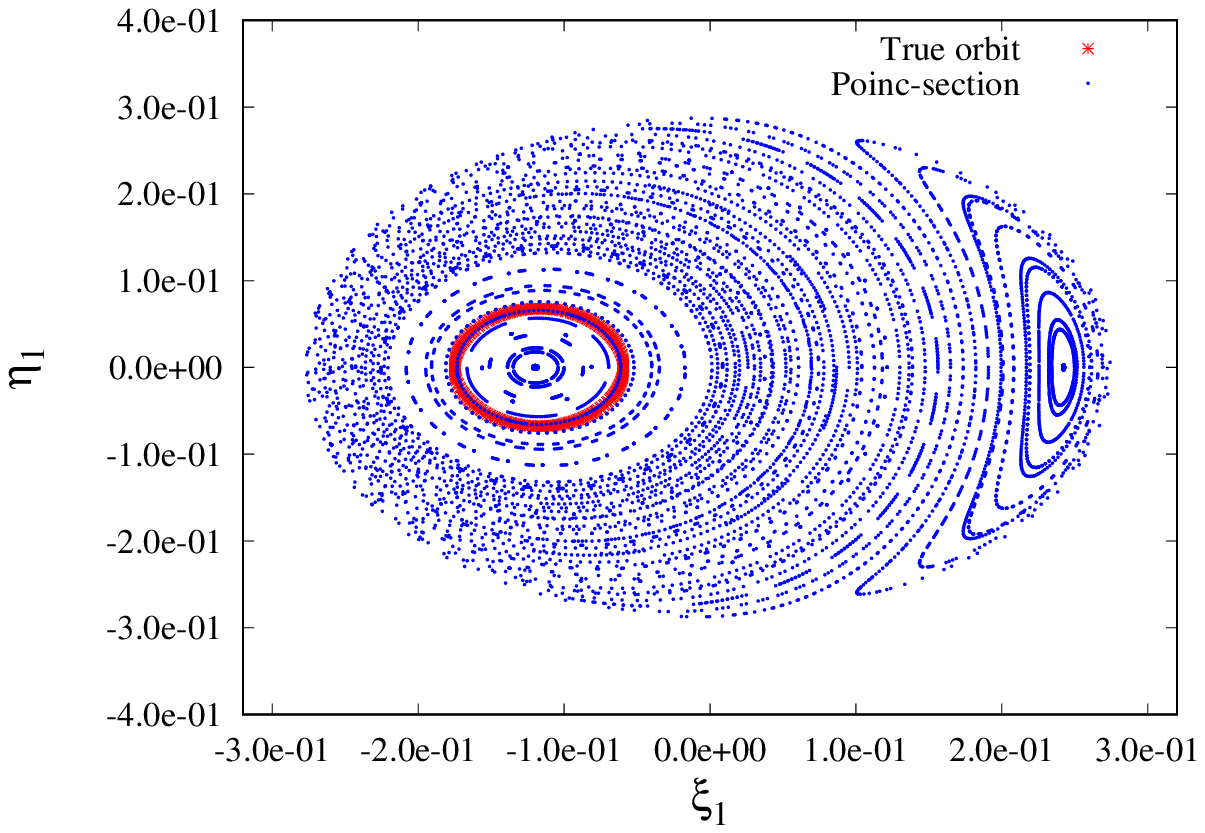}
  \includegraphics[width=6.0cm]{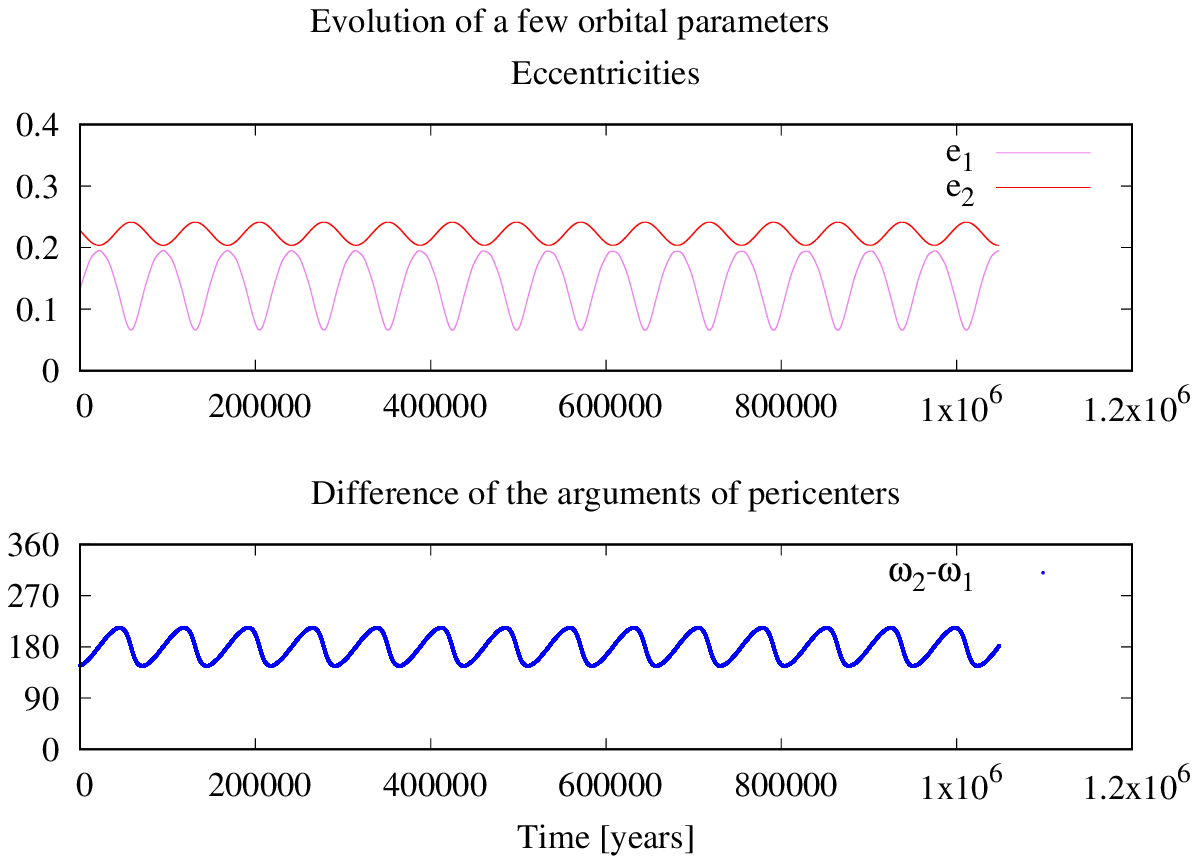}
  \caption{On the left, Poincar\'e sections that are corresponding to
    the hyperplane $\eta_2=0$ (with the additional condition
    $\xi_2>0$) and are generated by the flow of the Hamiltonian
    secular model $H^{({\rm sec})}$, which is given in~\eqref{frm:hsec}
    at order two in the masses for the exoplanetary system HD$\,$4732;
    the orbit in red refers to the motion starting from the initial
    conditions corresponding to the set $\Iscr_{4^{\circ}}$, that is
    described in~\eqref{def:set-cond-iniz}. On the right, evolution of
    secular orbital elements: the eccentricities of both the
    exoplanets (that are $e_1$ and $e_2$) and the difference of the
    arguments of the pericenters (i.e., $\omega_2-\omega_1$) are
    plotted on top and bottom, resp.}
  \label{fig:incl88-imut04-sezpoi}
\end{figure}

A few dynamical features of the Hamiltonian model defined by $H^{({\rm
    sec})}$ are summarized in the plots reported in
Figure~\ref{fig:incl88-imut04-sezpoi}. They refer, as an example, to
the initial conditions corresponding to the set of values
$\Iscr_{4^{\circ}}$, defined in~\eqref{def:set-cond-iniz}.  The
difference of the arguments of the pericenters $\omega_2-\omega_1$ is
plotted in the bottom-right panel of such a figure; then, we can
easily appreciate that this angle is librating around
$180^{\circ}$. By taking into account of the fact that the nodes are
opposite in the Laplace frame, this means that the pericenters of
HD$\,$4732b and HD$\,$4732c are in the so called ``apsidal locking''
regime in the vicinity of the alignment of the pericenters. This
phenomenon is expected to play a major role in making stable the
orbits for systems where the Keplerian part of the Hamiltonian is
strongly affected by the interactions between planets (see,
e.g.,~\cite{Mich-Mal-2004} or~\cite{Car-Loc-San-Vol-2021}). The
Poincar\'e sections of the motions starting from the initial
conditions corresponding to $\Iscr_{4^{\circ}}$ are plotted in red in
the panel on the left of Figure~\ref{fig:incl88-imut04-sezpoi} and it is
easy to remark that they are orbiting around a fixed point. Moreover,
it looks rather close to those sections marked in red, when their
distance from such a fixed point is compared with that from the orbits
that are enclosing another fixed point. Let us recall that all the
Poincar\'e sections reported in Figure~\ref{fig:incl88-imut04-sezpoi}
refer to the same level of energy, say $E$, corresponding to the set
of initial conditions $\Iscr_{4^{\circ}}$. Since $H^{({\rm sec})}$ is
a two degrees of freedom Hamiltonian, the manifold labeled by such a
value of the energy will be three-dimensional; in other words, by
plotting the Poincar\'e sections, we automatically reduce by one the
dimensions of the orbits. This is the reason why a fixed point
actually corresponds to a periodic orbit. Since the fixed point with
negative value of the abscissa is surrounded by closed curves, then we
can argue that such a periodic orbit is linearly stable for what
concerns the transverse dynamics. This means that it is a
one-dimensional elliptic torus, in the terminology we have adopted in
the present work. Therefore, we can conclude that the orbit generated
by the set $\Iscr_{4^{\circ}}$ of initial conditions is winding around
a linearly stable periodic orbit, by remaining in its vicinity.  This
explains why we are going to adopt a strategy based on two different
algorithms: the first one refers to the elliptic torus (that
corresponds to a fixed point in the Poincar\'e sections) and provides
a good enough approximation to start the second computational
procedure that constructs the final KAM torus (which shall include
also the points marked in red in
Figure~\ref{fig:incl88-imut04-sezpoi}).

\subsubsection{Explicit construction of the normal form for elliptic tori in the case of the secular model representing the planetary system HD$\,$4732.\\}\label{sss:semianal-4-ell-tori}
\noindent
The discussion above has highlighted that it is convenient to adopt a
suitable set of coordinates including also a resonant angle, that is
the difference of the arguments of the pericenters. In view of such a
target, we first introduce the set of action-angle variables
$(\vet\Jscr,\vet\psi)$ via the canonical transformation
\begin{equation}
\xi_j=\sqrt{2\Jscr_j}\cos \psi_j\ , \qquad \eta_j=\sqrt{2\Jscr_j}\sin \psi_j\ ,
\qquad \forall\ j=1,2,
\label{eq:azang}
\end{equation}
being $(\vet\xi,\vet\eta)$ the variables appearing as arguments of the
secular Hamiltonian $H^{({\rm sec})}$ defined in~\eqref{frm:hsec}. It
is important to recall that the angles $(\psi_1,\psi_2)$ associated to
these secular variables are nearly equal to the arguments of the
pericenters $(\omega_1,\omega_2)$, apart from a small correction due
to the transformation of coordinates induced by the application of the
Lie series $\exp \Lie{\chi_1^{(\Oscr2)}}$ to the Hamiltonian of the
three-body planetary problem.  Then, it is convenient to introduce a
new set of variables $(\vet I, \vet\theta)$ such that
\begin{equation}
\theta_1 = \psi_1 -\psi_2\ , \quad \theta_2 = \psi_2\ , \quad
I_1 = \Jscr_1\ ,  \quad I_2 = \Jscr_2 + \Jscr_1\ .
\label{eq:def-azang-Iphi}
\end{equation}
We now introduce the new canonical polynomial variables
$(\vet x, \vet y)$ defined as
\begin{equation}
x_j = \sqrt{2 I_j}\cos \theta_j\ , \qquad
y_j=\sqrt{2 I_j}\sin \theta_j\ , \qquad \forall\ j=1,2\ .
\label{eq:def-dalembcoord-xy}
\end{equation}
Let us also remark that making Poincar\'e sections with respect to the
hyperplane $\eta_2=0$, when $\xi_2>0$ is equivalent to impose
$\psi_2=0$, because of the definitions in~\eqref{eq:azang}. Therefore,
looking at
formul{\ae}~\eqref{eq:def-azang-Iphi}--\eqref{eq:def-dalembcoord-xy},
one can easily realize that the drawing in the left panel of
Figure~\ref{fig:incl88-imut04-sezpoi} can be seen as a plot of the
Poincar\'e sections in coordinates $(x_1\,,\,y_1)$ with respect to
$y_2=0$ and with the additional condition $x_2>0$. Revisiting the plot
in the bottom--right box of Figure~\ref{fig:incl88-imut04-sezpoi} in
the context of the new canonical variables is interesting, because it
makes clear that $\theta_1$ is librating around $180^\circ$. In fact,
we have that $\theta_1 = \psi_1 -\psi_2 \simeq \omega_1 -\omega_2\,$,
because the relation between these differences of angles is given by
the transformation induced by the application of the Lie series $\exp
\Lie{\chi_1^{(\Oscr2)}}$, that is close to the identity.

By a numerical method\footnote{Let us imagine to start from an initial
  condition denoted by $( \hat{\vet x}, \hat{\vet y})$ that is close
  enough to the periodic orbit generated by the wanted solution $(
  \vet x^\star, \vet y^\star)$; typically, at the beginning one can
  put $( \hat{\vet x}, \hat{\vet y})$ equal to the values assumed by
  the canonical variables $( \vet x, \vet y)$ in correspondence with
  the set $\Iscr_{i_{\rm mut}(0)}$, defined
  in~\eqref{def:set-cond-iniz}. During a long enough numerical
  integration of the Hamilton equations related to $H^{({\rm sec})}$,
  one can easily determine $\hat x_{1,-}$ and $\hat x_{1,+}$ that are
  the minimum value assumed by the variable $x_1$ in correspondence
  with the Poincar\'e sections and the maximum one, resp.  If the
  difference $\hat x_{1,+}-\hat x_{1,-}$ is below a prescribed (small)
  threshold of tolerance, then we assume to know the solution with a
  good enough level of approximation and we stop this computational
  procedure by setting $( \vet x^\star, \vet y^\star)=( \hat{\vet x},
  \hat{\vet y})$. If such a ``way out condition'' is not satisfied,
  then we define $x_1^\star=(\hat x_{1,+}+\hat x_{1,-})/2$,
  $y_1^\star=0$, $y_2^\star=0$ and we determine the positive value of
  $x_2^\star$ so that the energy level of this new approximation of
  the final solution, i.e., $( \vet x^\star, \vet y^\star)$, is still
  equal to the value $E$ corresponding to the set $\Iscr_{i_{\rm
      mut}(0)}$. Let us remark that in the (re)definition of $( \vet
  x^\star, \vet y^\star)$ we are exploiting both the definition of the
  Poincar\'e sections and their symmetry with respect to the axis of
  the abscissas. At this point, we put $( \hat{\vet x}, \hat{\vet
    y})=( \vet x^\star, \vet y^\star)$ and we restart the
  computational procedure by performing another numerical integration
  so to determine new values of $\hat x_{1,-}$ and $\hat x_{1,+}$ and
  so on, until the ``way out condition'' will be satisfied.}, we can
easily determine the initial condition $( \vet x^\star, \vet y^\star)$
that is in correspondence with a Poincar\'e section and generates a
periodic solution. We can now subdivide the variables in two different
couples. The first one is given by $(p,q)\in \reali\times \toro$,
i.e., the action-angle couple describing the periodic motion.  Thus,
we rename the angle $\phi_2$ as $q$, while the action is obtained by
translating the origin of $I_2$ so that $p= I_2 -I^\star$, where at
the first trial\footnote{See the discussion about the solution of the
  implicit equation~\eqref{frm:Newt-En-sez-Poi} by using the Newton
  method, which is reported at the end of these explanations.} the
shift value $I^\star$ is fixed so that $I^\star=
\big((x_2^{\star})^2+(y_2^{\star})^2\big)/2$. For what concerns the
second couple of canonical coordinates, we start from the polynomial
variables $(x_1,y_1)$ in order to describe the motion transverse to
the periodic orbit. The last preliminary translation is on $x_1\,$, in
order to have expansions around the value $x_1^\star$, given by the
initial condition computed numerically.  Let us emphasize that, since
the fixed point we are trying to approximate in
Figure~\ref{fig:incl88-imut04-sezpoi} corresponds to
$\phi_1=180^{\circ}$, we have that $y_1^{\star}= 0$ and here a
translation is not needed. It is now convenient to rescale the
transverse variables $(\bar x_1,y_1)$, being $\bar x_1= x_1 -
x_1^\star$, in such a way that the Hamiltonian part which is quadratic
in the new variables $(x,y)$ and does not depend on $(p,q)$ is in the
form $\Omega^{(0)}(x^2 + y^2)/2$. This rescaling can be done by a
canonical transformation as the quadratic part does not have any mixed
term $\bar x_1 y_1$ and the coefficients of $\bar x_1^2$ and $y_1^2$
have the same sign, because of the proximity to an elliptic
equilibrium point. Thus, since such a quadratic part is in the
preliminary form $a\bar x_1^2+ b y_1^2$, it suffices to define the new
variables $(x,y)$ as $x = \sqrt[4]{\frac a b}\, \bar x_1, \ y =
\sqrt[4]{\frac b a}\, y_1\,$. Finally, we introduce the second pair of
canonical coordinates $(J,\phi)\in \reali_+\cup\{0\}\times \toro$ so
that $x=\sqrt{2 J}\cos\phi$ and $y=\sqrt{2 J}\sin\phi$.

In the case of the secular dynamics of the planetary system
HD$\,$4732, starting from $H^{({\rm sec})}$ in~\eqref{frm:hsec}, we
have applied all the canonical transformations listed above and we
have expanded the Hamiltonian $\Hscr^{(0)}(p,q,J,\phi)$ up to
degree~$16$ in the square roots of the actions $(p,J)$. Since
$\Hscr^{(0)}(p,q,J,\phi)$ is in a suitable form to apply the algorithm
fully described in Subsection~\ref{sbs:Kolmog-4-ell-tori} in the case
with $n_1=n_2=1$ (this is the reason why all the variables
$(p,q,J,\phi)$ are here denoted as scalar quantities instead of
vectorial ones), we have applied such a computational procedure. We
have performed~$19$ steps of the normalization algorithm so producing
$\Hscr^{(19)}(p,q,J,\phi)$. During those computations, the Fourier
expansions in $q$ of all the Hamiltonians defined by the algorithm
have been truncated at a maximal trigonometric degree equal to $40$;
since $K=2$, this choice allows to properly determine the generating
functions for the first~$20$ normalization steps. For the sake of
brevity, we omit to report the graphs of the norms of all the
generating functions that are defined by the normalization procedure,
also because those plots are similar to the corresponding ones
included in~\cite{Car-Loc-2021}
and~\cite{Car-Loc-San-Vol-2021}. Indeed, they show that the
convergence to the identity of the canonical transformations defined
at the $r$-th step of the algorithm is very fast with respect to $r$.
This fact also allows to iterate a few times all the normalization
procedure constructing the normal form for an elliptic torus with a
computational cost which is not too expensive. We are interested in
doing that in order to refine the choice of the initial shift value
$I^\star$. Since all other canonical transformations are unambigously
defined, we have some remaining arbitrariness just on the translation
$p= I_2 -I^\star$. We finally determine $I^\star$ in such a way that
\begin{equation}
  \label{frm:Newt-En-sez-Poi}
  \Escr^{(19)}\big(I^\star\big) = E\ ,
\end{equation}
where $E$ is the energy level of the Poincar\'e sections and
$\Escr^{(19)}\big(I^\star\big)$ is the energy of the elliptic torus in
the approximation provided after~$19$ steps of normalization.  The
implicit equation above can be numerically solved in the unknown
$I^\star$ by iterating a few times the Newton method; this is done
starting from the initial guess
$\big((x_2^{\star})^2+(y_2^{\star})^2\big)/2$, according with the
discussion above.

For brevity, we omit also the tests showing that there is an excellent
agreement between the wanted periodic orbit and the nearly invariant
curve, which is provided by the last execution of the normalization
algorithm, that is launched during the final iteration of the Newton
method targeting the solution
of~\eqref{frm:Newt-En-sez-Poi}. Actually, it corresponds to the
counter-image of the set $(p=0,\,q\in\toro,\,J=0,\,\phi=0)$ and is
expressed in the coordinates $(\vet\xi,\vet\eta)$, after having
composed all the previous canonical transformations.

\subsubsection{Explicit construction of the normal form for KAM tori in the case of the secular model representing the planetary system HD$\,$4732.\\}\label{sss:semianal-4-KAM-tori}
Since the Hamiltonian $\Hscr^{(19)}(p,q,J,\phi)$ is very close to the
normal form related to the wanted elliptic torus, we use it as the
starting point to construct a semi-analytic solution that should
provide a good approximation of the orbits generated by the initial
conditions corresponding to the set $\Iscr_{4^{\circ}}\,$. For such a
purpose, first we translate once again the coordinates. This is made
in such a way that the new invariant torus we are going to construct
will be located in the proximity of these initial conditions;
therefore, we define two new pairs of action-angle coordinates $(\vet
p,\vet q)\in \reali^2\times \toro^2$. It is convenient to set
$p_2=J-J^{\star}$, being $J^{\star}$ the value of the momentum $J$
computed in correspondence with the initial conditions related to the
set $\Iscr_{4^{\circ}}\,$, that generate the Poincar\'e sections
marked in red in Figure~\ref{fig:incl88-imut04-sezpoi}. We also
introduce $p_1=p-p^{\star}$, with
$p^{\star}=-\big(\Omega^{(19)}/\omega^{(19)}\big)J^{\star}$, being
$2\pi/\omega^{(19)}$ approximately equal to the period of the motion
on the previously determined one-dimensional elliptic torus, while the
angular velocity of the transverse (small) oscillations in its
vicinity is close to $\omega^{(19)}$. We recall that the values of
both $\omega^{(19)}$ and $\Omega^{(19)}$ appear in the
expansion~\eqref{frm:espansione-H^(r)} of the Hamiltonian
$\Hscr^{(19)}$, that is provided at the end of the previous
normalization algorithm. Moreover, we rename the angles $(q,\phi)$ as
$(q_1,q_2)$, respectively; then, we perform the two translations
described just above, by expanding the new Hamiltonian $H^{(0)}(\vet
p,\vet q)$ up to degree~$8$ in the actions $\vet p$. By considering
just the integrable approximations of $\Hscr^{(19)}$ and $H^{(0)}$
(this means that the terms depending by the angles are temporarily
neglected), one can easily realize that the energy constant $E^{(0)}$
corresponding to the new Hamiltonian is such that $E^{(0)}\simeq E$,
because of the equation
$\omega^{(19)}p^{\star}+\Omega^{(19)}J^{\star}=0$ that is due to the
definitions of the shift values $\big(p^{\star},J^{\star}\big)$.
Since $H^{(0)}(\vet p,\vet q)$ is in a suitable form to apply the
algorithm fully described in Subsection~\ref{sbs:Kolmog-original}, we
have performed~$19$ steps of such a computational procedure too, so
producing $H^{(19)}(\vet p,\vet q)$. During these computations, the
Fourier expansions in $q$ of all the Hamiltonians defined by the
normalization algorithm have been truncated at a maximal trigonometric
degree equal to $40$. This choice allows to properly determine the
generating functions $\chi_1^{(r)}$ and $\chi_2^{(r)}$ for the
first~$20$ normalization steps.

\begin{figure}
\centering
  \includegraphics[width=6.0cm]{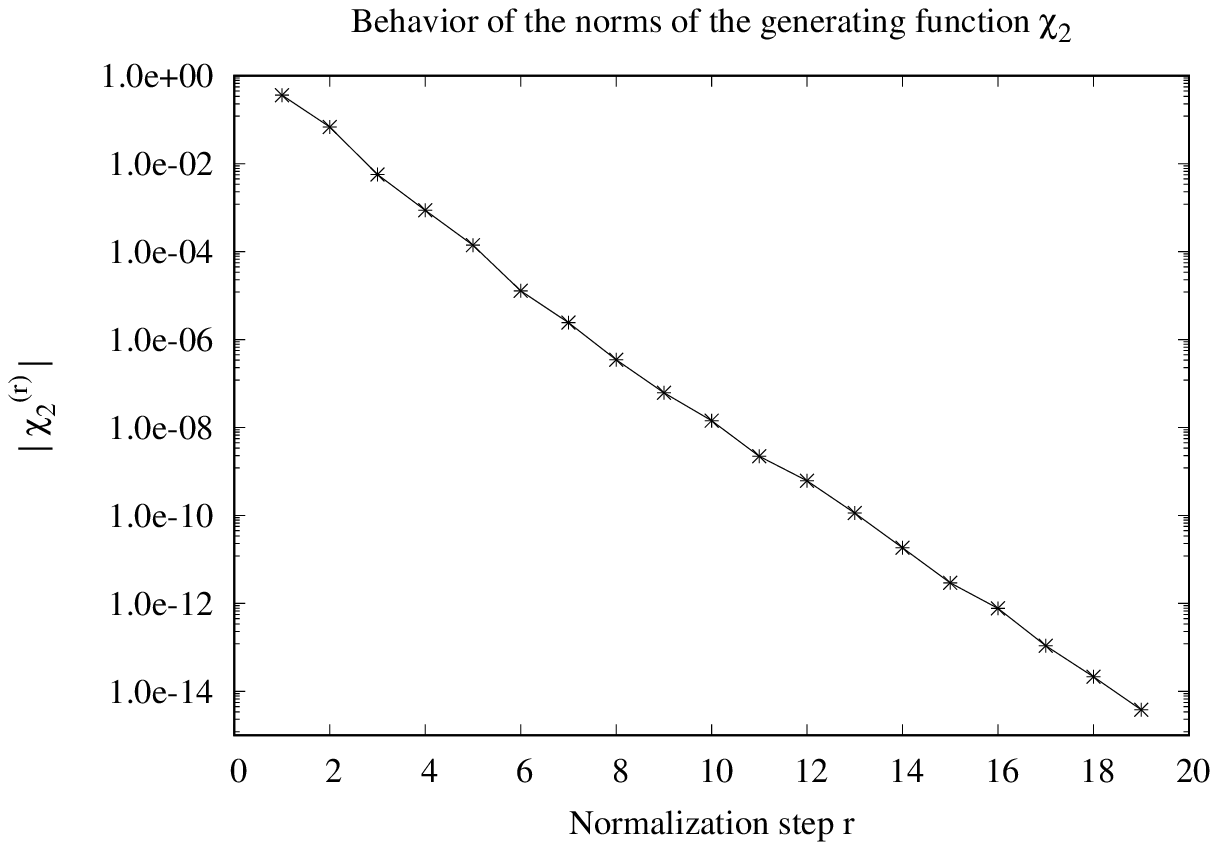}
  \includegraphics[width=6.0cm]{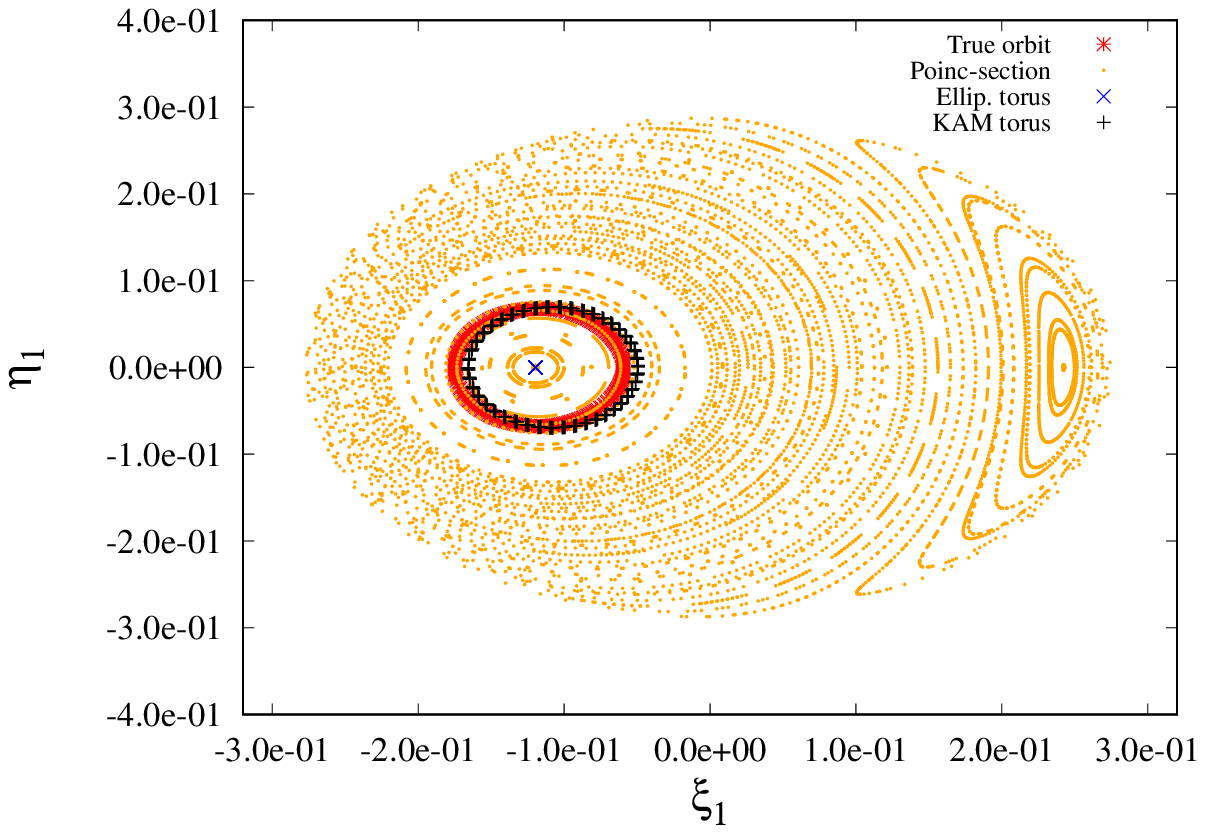}
  \caption{On the left, study of the decrease of
    $\big\|\chi_2^{(r)}\big\|$ as a function of the normalization step
    $r$. On the right, comparisons between the Poincar\'e sections
    generated by two different initial conditions, that correspond to
    the set $\Iscr_{4^{\circ}}$ and a point on the (nearly) invariant
    torus $\vet p=\vet 0$ related to the Hamiltonian $H^{(19)}$,
    respectively. The former ones are marked in red as in the left
    panel of Figure~\ref{fig:incl88-imut04-sezpoi}, while the latter
    ones are in black. The Poincar\'e sections are defined in the same
    way as those reported in Figure~\ref{fig:incl88-imut04-sezpoi}; in
    particular, the dots plotted in blue there are located exactly in
    the same positions as those marked in orange here. The blue symbol
    $\times$ refers to the motion on the elliptic torus corresponding
    to the Hamiltonian $\Hscr^{(19)}$.  }
  \label{fig:incl88-imut04-KAM}
\end{figure}

\noindent
It is convenient to define the norms of the generating functions as
the sum of the absolute values of the coefficients appearing in their
(finite) Taylor-Fourier expansions.  In the left panel of
Figure~\ref{fig:incl88-imut04-KAM}, we report the plot of
$\big\|\chi_2^{(r)}\big\|$ in a semi-log scale and as a function of
the normalization step $r$, while we have decided to not include also
$\big\|\chi_1^{(r)}\big\|$, because for every $r$ it is definitely
smaller than $\big\|\chi_2^{(r)}\big\|$. One can appreciate that the
geometrical decrease of the generating functions is very sharp and
regular; therefore, this shows that the normalization algorithm
constructing the Kolmogorov normal form is convergent in a quite
rapid way.

We can now check the quality of our results. Let us denote with
$\Cscr$ the canonical transformation we obtain by composing all the
changes of coordinates we have discussed in the present
Subsection~\ref{sec:normal-form-4-tori}.  Therefore, we have that
$(\vet{\xi},\vet{\eta})=\Cscr(\vet{p},\vet{q})$, where
$(\vet{\xi},\vet{\eta})$ are the canonical coordinates referring to
the Hamiltonian secular model $H^{({\rm sec})}$, that is defined
in~\eqref{frm:hsec}, while $(\vet{p},\vet{q})$ are the action-angle
variables that are introduced at the end of the previously described
computational procedure.  Inspired by the semi-analytic
scheme~\eqref{frm:semi-anal-scheme}, which provides a way to integrate
the Hamilton equations, we start by computing
$(\vet{\xi}(0),\vet{\eta}(0))=\Cscr(\vet{0},\vet{0})$. Since
$H^{(19)}$ is very close to be in Kolmogorov normal form and $H^{({\rm
    sec})}\simeq H^{(19)}\big(\Cscr(\vet{p},\vet{q})\big)$ (the
discrepancies are mainly due to the unavoidable truncations that are
made on the expansions of the Hamiltonians), then
$\big(\vet{\xi}(t),\vet{\eta}(t)\big)=\big(\Cscr(\vet{0},\vet{\omega}^{(19)}t)\big)$
provides a good approximation of the flow induced by $H^{({\rm
    sec})}$.  We also recall that the values of the angular velocity
vector $\vet{\omega}^{(19)}$ appear in the
expansion~\eqref{frm:espansione-HKAM^(r)} of the Hamiltonian
$H^{(19)}$.  Computing the Poincar\'e sections of the motion law
$\big(\Cscr(\vet{0},\vet{\omega}^{(19)}t)\big)$ is not very
comfortable; therefore, it is convenient to refer to its approximation
which is given by the numerical solution of the Hamilton equations for
$H^{({\rm sec})}$ starting form the initial conditions
$(\vet{\xi}(0),\vet{\eta}(0))=\Cscr(\vet{0},\vet{0})$. The Poincar\'e
sections we have obtained in this way are plotted in black on the
right panel of Figure~\ref{fig:incl88-imut04-KAM}. They are in good
agreement with the the Poincar\'e sections marked in red in both
figures~\ref{fig:incl88-imut04-sezpoi}
and~\ref{fig:incl88-imut04-KAM}, that refer again to the flow induced
by $H^{({\rm sec})}$, but starting from the initial conditions related
to the set $\Iscr_{4^{\circ}}\,$.  This confirms that we are able to
obtain reliable approximations of the secular motions for extrasolar
planetary systems, by using computational procedures based on the
construction of suitable (Kolmogorov-like) normal forms.

\subsubsection{Final comments about our semi-analytic results.\\}\label{sss:final-comments}
Looking closely at the right panel of
Figure~\ref{fig:incl88-imut04-KAM}, one can observe that the
Poincar\'e sections plotted in black goes from the part internal to
the orbit in red to the external one and vice versa.  This provides a
clear indication that the energy level of the final KAM torus (that is
$\simeq E^{(19)}$) is not very close to that of all the Poincar\'e
sections plotted in Figure~\ref{fig:incl88-imut04-sezpoi} (being $E$
its value). Indeed, the relative error $\big|E^{(19)}-E\big|/|E|$ is
about $12\,$\%. The agreement between the results produced by the
purely numerical integrations or by adopting our semi-analytical
approach can be strongly improved by a suitable further refinement of
our computational procedure. The description of such an extension goes
beyond the scopes of the present work, but we stress that it can be
done so as to ensure also that the condition on the coherence with the
energy of the Poincar\'e sections, i.e.,
\begin{equation}
  \label{frm:Newt-KAM-En-sez-Poi}
  E^{(R_{\rm I})} = E\ ,
\end{equation}
is satisfied within a tolerance range that is acceptable for a
numerical solution of the equation above, where $R_{\rm I}$ is the
number of steps that are explicitly performed in order to construct
the final Kolmogorov normal form. Here, we limit ourselves to
anticipate some of the results that can be obtained by implementing
that further refinement, in order to let the reader appreciate the
power of this kind of methods. For what concerns the planetary system
HD$\,$4732 we already have studied the motions starting from the
following sets of initial conditions: $\Iscr_{2^{\circ}}\,$,
$\Iscr_{4^{\circ}}\,$, $\Iscr_{6^{\circ}}\,$, $\ldots$
$\Iscr_{40^{\circ}}\,$. We can construct invariant KAM tori well
approximating the orbits for all these cases, except those
corresponding to the sets $\Iscr_{32^{\circ}}$ and
$\Iscr_{34^{\circ}}$. We emphasize that these limitations are due to
real dynamical phenomena. The Poincar\'e sections generated by those
initial conditions clearly shows that between $34^{\circ}$ and
$36^{\circ}$ there is the transition from the librations to the
circulation regime, for what concerns the difference of the argument
of the pericenters. Moreover, this kind of orbits are observed in
stable situations up to initial values of the mutual inclinations that
are about $40^{\circ}$, while for even larger angles there are robust
configurations just inside the Lidov-Kozai resonance, which has
different dynamical features (see~\cite{Vol-Roi-Lib-2019}).  As we
have already mentioned above, we plan to describe these new results in
a forthcoming work.

The evolution of the eccentricities plotted in the right panel of
Figure~\ref{fig:incl88-imut04-sezpoi} clearly shows that their average
value is larger than~$0.1$ for both the exoplanets orbiting around
HD$\,$4732. Therefore, the new approach that we have introduced in the
present work behaves definitely better with respect to the previous
one, which was described in~\cite{Vol-Loc-San-2018} and was shown to
be successful just for systems with exoplanetary eccentricities
smaller than $0.1\,$.  In our opinion the main source of improvement
is due to the new strategy, because it combines the preliminary
construction of the normal form for a suitable elliptic torus with the
final one, which is performed in its vicinity for a KAM torus whose
shape is a good approximation of the secular orbits. In order to
mention another relevant success of our new approach, let us stress
that in~\cite{Car-Loc-San-Vol-2021} we applied it also to
the delicate case of a system including both the two largest
exoplanets orbiting around $\upsilon$~Andromed{\ae}~A and the star
itself.

\vskip 0.4 cm
\noindent
{\bf Acknowledgements.} This work was partially supported by the
project MIUR-PRIN 20178CJA2B ``New frontiers of Celestial Mechanics:
theory and applications''. M.V. thanks the ASI Contract
n. 2018-25-HH.0 (Scientific Activities for JUICE, C/D phase).
Moreover, we extend our gratitude also to the MIUR Excellence
Department Project awarded to the Department of Mathematics of the
University of Rome ``Tor Vergata'' (CUP E83C18000100006), which made
available the computational resources we exploited.

%
%

\end{document}